\documentclass{report}
\usepackage{amssymb,a4wide}
\usepackage{epsfig}

\newcommand{\complex}{\hbox{$\mathbb C$}}
\newcommand{\zet}{\hbox{$\mathbb Z$}}
\newcommand{\reals}{\hbox{$\mathbb R$}}

\def\bra#1{\langle #1 \vert}
\def\ket#1{\vert #1 \rangle}
\def\be{\begin{equation}}
\def\ee{\end{equation}}
\def\ba{\begin{eqnarray}}
\def\ea{\end{eqnarray}}
\newcommand{\nn}{\nonumber \\}
\newcommand{\no}{\nonumber}
\newcommand{\pn}{\par\noindent}
\def\A{{\cal A}}
\def\D{{\cal D}}
\def\L{{\cal L}}

\def\hh{{1 \over 2}}
\def\hq{\frac{1}{q}}

\begin{document}

\begin{center}
{\Large \bf $q$-deformed Heisenberg Algebras}
\footnote{Lectures given at the 38. Internationale Universit\"atswochen f\"ur Kern-
und Teilchenphysik, Schladming, Styria, Austria, January 9 - 16, 1999}\\[3ex]
J.WESS\\[3ex]
{\it Sektion Physik der Ludwig-Maximilians-Universit\"at\\
Theresienstr. 37, D-80333 M\"unchen\\
and\\
Max-Planck-Institut f\"ur Physik\\
(Werner-Heisenberg-Institut)\\
F\"ohringer Ring 6, D-80805 M\"unchen\\
}
\end{center}
\vskip2cm
\thispagestyle{empty}
\tableofcontents

\newpage
\section*{Introduction}
This lecture consists of two sections. In section 1 we consider the simplest
version of a $q$-deformed Heisenberg algebra as an example of a 
noncommutative structure. We first derive a calculus entirely based on 
the algebra and then formulate laws of physics based on this calculus.
Then we realize that an interpretation of these laws is only possible if
we study representations of the algebra and adopt the quantum mechanical
scheme. It turns out that observables like position or momentum have
discrete eigenvalues and thus space gets a lattice-like structure.

In section 2 we study a framework for higher dimensional noncommutative
spaces based on quantum groups. The Poincar{\'e}-Birkhoff-Witt property and
conjugation properties play an essential role there. In these spaces
derivatives are introduced and based on these derivatives a
$q$-deformed Heisenberg algebra can be constructed.




\newpage
\chapter{$q$-Deformed Heisenberg algebra in one dimension}\label{Sekt1}
\section{A calculus based on an algebra}\label{sub1}

We try to develop a formal calculus entirely based on an algebra. In 
this lecture we consider the q-deformed Heisenberg algebra as an
example and define derivatives and an integral on purely 
algebraic grounds. The functions which we differentiate and integrate
are elements of a subalgebra - we shall call them fields. The integral
is the inverse image of the derivative - thus it is an indefinite
integral. For the derivatives a Leibniz rule can be found quite 
naturally and this leads to a rule for partial integration.
\bigskip
\pn
\underline{The algebra}

As a model for the noncommutative structure that 
arises from quantum group considerations we start from 
the q-deformed relations of a Heisenberg algebra.

\ba
& q^{\frac{1}{2}} xp-q^{-\frac{1}{2}} px = i \Lambda, \nn
& \Lambda p=q p \Lambda, \quad \Lambda x=q^{-1} x \Lambda \label{1.1} \\
& q \in \reals, \quad q \neq 0 \no
\ea

As it will be explained in the following the algebra has to be 
a star algebra to allow a physical interpretation. We are now going
to argue for such an algebra that is based on the relations
(\ref{1.1}).
The element $x$ of the algebra will be identified with the
observable for position in space, the element $p$ with the
canonical conjugate observable, usually called momentum. 
Observables have to be represented by selfadjoint
linear operators in a Hilbert space. This will guarantee real
eigenvalues and a complete set of orthogonal eigenvectors.

For this reason we require already at the level of the algebra
an antilinear involution that will be identified with the
conjugation operation of linear operators in Hilbert space.
At the algebraic level we use bar and at the operator level
star to denote this involution. Algebraic selfadjoint elements will
have to be represented by selfadjoint operators in Hilbert space.

As a first step we extend the algebra (\ref{1.1}) by
conjugate elements $\overline{x}$, $\overline{p}$, $\overline{\Lambda}$
and find from (\ref{1.1}) ($\overline{q}=q$):

\be
q^{\frac{1}{2}} \overline{p} \,\overline{x} -q^{-\frac{1}{2}} \overline{x}
\, \overline{p}  = -i \overline{\Lambda}
\label{1.2}
\ee
Then for reasons explained above we demand:
\be
\overline{x}=x, \quad \overline{p}=p
\label{1.3}
\ee
With these conditions follows from (\ref{1.2}) 
and (\ref{1.1})   for $q \neq 1$:

\ba
px &=& i \lambda^{-1} (q^{-\frac{1}{2}} \Lambda-q^{\frac{1}{2}}
\overline{\Lambda}) 
\label{1.4} \\
xp &=& i \lambda^{-1} (q^{\frac{1}{2}} \Lambda-q^{-\frac{1}{2}}
\overline{\Lambda}), \quad \lambda=q-\frac{1}{q} \no
\ea

For $\Lambda$ selfadjoint this leads to $xp+px=0$.
This is too strong a relation which we do not admit as it does not allow
a smooth transition for $q \rightarrow 1$. In great generality we can
demand $\Lambda$ to be the product of a unitary and a hermitean element.
The simplest choice is for $\Lambda$ to be unitary. Thus we extend the
algebra by $\Lambda^{-1}$ and demand

\be
\overline{\Lambda}=\Lambda^{-1}
\label{1.5}
\ee

We want to have $x^{-1}$ as an observable as well, thus we extend the 
algebra once more, this time by $x^{-1}$.

The following ordered monomials form a basis of the algebra:

\be
x^m \Lambda^n, \quad m,n \in \zet
\label{1.6}
\ee

The element $p$ can be expressed in this basis 
\be
p = i \lambda^{-1} x^{-1} (q^{\frac{1}{2}} \Lambda-q^{-\frac{1}{2}}
\overline{\Lambda})
\label{1.7}
\ee
This follows from (\ref{1.4}).

To summarize, we study the associative algebra 
over the complex numbers freely generated by the
elements $p$, $x$, $\Lambda,x^{-1}$, $\Lambda^{-1}$ and
their conjugates. This algebra is to be divided by the ideal generated by 
the relations (\ref{1.1}),(\ref{1.3}) and (\ref{1.5})
and those that follow for $x^{-1}$ and $\Lambda^{-1}$. 

\bigskip
\pn
\underline{Fields and derivaties:}

At the algebraic
level we define a field $f$ as an element of the subalgebra
generated by $x$ and $x^{-1}$, then completed
by formal power series.

\be
f(x) \in \Big[ [ x,x^{-1} ] \Big] \equiv {\cal{A}}_x
\label{1.8}
\ee

A derivative we define as a map of ${\cal{A}}_x$ into ${\cal A}_x$, as 
we are going to explain now.
From the algebra follows:

\be
p f(x)=g(x) p-iq^{\frac{1}{2}} h(x) \Lambda
\label{1.9}
\ee
where $g(x)$ and $h(x)$ can be computed 
using (\ref{1.1}). The derivative is  defined as follows:

\be
\nabla: {\cal A}_x \rightarrow {\cal A}_x, \quad \nabla f(x)=h(x)
\label{1.10}
\ee

The monomials $x^m$, $m \in \zet$ form a basis of ${\cal A}_x$.
On these elements the derivative acts as follows:

\be
\nabla x^m=[m] x^{m-1}, \quad [m]=\frac{q^m-q^{-m}}{q-q^{-1}}
\label{1.11}
\ee

We see that the element $x^{-1}$ 
is not in the image of $\nabla$. The $\nabla$ map also has a kernel,
the constants:

\be
\nabla c =0, \quad c \in \complex
\label{1.12}
\ee

In a similar way we can define the maps $L$ and $L^{-1}$ from
${\cal A}_x$ onto ${\cal A}_x$. We start from the algebraic relation

\be
\Lambda f(x)=j(x) \Lambda, \quad \Lambda^{-1} f(x)=k(x) \Lambda^{-1}
\label{1.13}
\ee
and define

\ba
L: {\cal A}_x \rightarrow  {\cal A}_x, && L f(x)=j(x) 
\label{1.14} \\
L^{-1}: {\cal A}_x \rightarrow  {\cal A}_x, && L^{-1} f(x)=k(x) \no
\ea
For the $x$-basis we obtain:

\be
L x^m=q^{-m} x^m, \quad L^{-1} x^m=q^m x^m
\label{1.15}
\ee

The elements $x$, $x^{-1}$ of the algebra ${\cal A}_x$ define a map
${\cal A}_x \rightarrow  {\cal A}_x$ in a natural way.

These maps form an algebra
\ba
& L x=q^{-1} x L, \quad L \nabla =q \nabla L,
\label{1.16} \\
& q^{\frac{1}{2}} x \nabla-q^{-\frac{1}{2}} \nabla x=-q^{-\frac{1}{2}} L \no
\ea
homomorphic to the algebra (\ref{1.1})
with the identification:
\be
L \sim \Lambda, \quad x \sim x, \quad -i q^{\frac{1}{2}} \nabla \sim p
\label{1.17}
\ee

We do not consider the complex extension of the algebra (\ref{1.16})
and do  not define a bar operation on $L$ and $\nabla$. 
Nevertheless, it can be verified directly from the definition 
of $L,L^{-1}$ and $\nabla$ that

\ba
\nabla&=&\lambda^{-1} x^{-1}(L^{-1}-L) \label{1.18} \\
\nabla x^m&=&\frac{1}{\lambda} (q^m-q^{-m}) x^{m-1}=[m] x^{m-1}
\no
\ea
which agrees with (\ref{1.11}).

\bigskip
\pn
\underline{Leibniz rule:}

For usual functions we know the Leibniz rule:
\be
\partial fg=(\partial f) g+ f(\partial g)
\label{1.19}
\ee

There is a Leibniz rule for the derivative $\nabla$ 
as well. It can  be obtained from (\ref{1.18}) 
if we know how $L$ and $L^{-1}$ acts on the product of fields.
We compute this action by taking products of  elements
in the $x$-basis:

\ba
L x^n x^m&=&L x^{m+n}=q^{-(m+n)} x^{m+n} \label{1.20} \\
&=&q^{-m} x^m q^{-n} x^n =(L x^m) (Lx^n) \no
\ea
Similar for $L^{-1}$:

\be
L^{-1} x^n x^m =(L^{-1} x^m) (L^{-1} x^n)
\label{1.21}
\ee
This leads to the rule for the product of arbitrary elements
of ${\cal A}_x$:

\ba
L fg= (L f) (L g) \label{1.22} \\
L^{-1} fg= (L^{-1}  f) (L^{-1}g) \no
\ea

For the maps $x$, $x^{-1}$ we have the obvious rule:

\ba
x \; fg=(xf)g \equiv f(xg) \label{1.23} \\
x^{-1} \; fg=(x^{-1} f)g \equiv f(x^{-1} g) \no
\ea
These formulas can be used to obtain the Leibniz rule
for $\nabla$, using (\ref{1.18}):

\ba
\nabla \; fg&=&\lambda^{-1} x^{-1} (L^{-1} -L) \; fg
\label{1.24} \\
&=& \lambda^{-1} x^{-1} \Big( (L^{-1} f) ( (L^{-1} g) - (Lf)(Lg) \Big) 
\no
\ea

Now we form a derivative on $f$ for the first and on $g$ 
for the second term:

\ba
\nabla \; fg&=&\Big(\lambda^{-1} x^{-1} (L^{-1} -L) f \Big)(L^{-1} g) 
+ \lambda^{-1} (x^{-1} L f)(L^{-1}g) \label{1.25}\\
&&+ (Lf) \lambda^{-1} x^{-1} (L^{-1} -L) g
-\lambda^{-1} (L f)(x^{-1} L^{-1}) g
\no
\ea
The result is:
\be
\nabla \; fg =(\nabla f)(L^{-1} g) +(Lf)(\nabla g)
\label{1.26}
\ee

The role of $f$ and $g$ can be exchanged, $f$ and $g$
commute. In a similar way that  led from (\ref{1.24}) to
(\ref{1.25}) we could have formed the derivative on
$g$ for the first term and on $f$ for the second term of
(\ref{1.24}). The result then is:

\be
\nabla \; fg =(\nabla f)(L g) +(L^{-1} f)(\nabla g)
\label{1.27}
\ee
The expressions (\ref{1.26}) and (\ref{1.27})
are identical due to the identity in (\ref{1.23}).

There is also a $q$-version of Green' s theorem. We compute:

\ba
\nabla (\nabla f) (L^{-1} g)=(\nabla^2 f) g+(L^{-1} \nabla f)(\nabla L^{-1} g)
\label{1.28} \\
\nabla (L^{-1} f) (\nabla  g)=(\nabla L^{-1} f) (L^{-1} \nabla g)+
f (\nabla^2 g)
\no
\ea

The two versions of the Leibniz rule (\ref{1.26}) and 
(\ref{1.27}) have been used. We subtract the two equations and obtain
Green' s theorem:

\be
(\nabla^2 f)(g)-(f)(\nabla^2 g)=\nabla \Big( (\nabla f)(L^{-1} g)-
(L^{-1} f)(\nabla g)\Big)
\label{1.29}
\ee

\bigskip
\pn
\underline{The indefinite integral:}

We define the indefinite integral over a field as the
inverse image of the derivative (\ref{1.10}) and (\ref{1.11}).
We know that $x^{-1}$ is not in the range of $\nabla$
and that the constants are in the kernel of $\nabla$.

\be
\int^x x^n=\frac{1}{[n+1]} x^{n+1}+c, \quad n \in \zet, \; n \neq -1
\label{1.30}
\ee

We can also use formulas (\ref{1.18}) to invert $\nabla$:

\be
\nabla^{-1}=\lambda \frac{1}{L^{-1}-L} x
\label{1.31}
\ee
This map is not defined on $x^{-1}$. We show that it reproduces
(\ref{1.30}) with $c=0$:

\ba
\nabla^{-1} x^n&=& \lambda \frac{1}{L^{-1}-L} x^{n+1} \nn
&=& \frac{\lambda}{q^{n+1}-q^{-n-1}} x^{n+1} \label{1.32} \\
&=& \frac{1}{[n+1]} x^{n+1} \no
\ea

The map $\nabla^{-1}$ is now defined on any field that when expanded in the
$x^n$ basis does not have an $x^{-1}$ term.

\ba
\nabla^{-1} f(x) &=& \lambda \sum_{\nu=0}^{\infty} L^{2 \nu} L x f(x)
\label{1.33} \\
&=& - \lambda \sum_{\nu=0}^{\infty} L^{-2 \nu} L^{-1} x f(x) \no
\ea

We shall use the first or second expansion depending on what  
series converges. As an example:

For $n\geq0$

\ba
\nabla^{-1} x^n&=& \lambda \sum_{\nu=0}^{\infty} L^{2 \nu} L x^{n+1} 
\label{1.34} \\
&=&  \lambda \sum_{\nu=0}^{\infty} q^{-(2 \nu+1)(n+1)} x^{n+1} \no
\ea
This sum converges for $q>1$ and we obtain (\ref{1.32}) from 
(\ref{1.34}).

For $n<-1$ we find that the second expansion of (\ref{1.33}) 
converges and gives again the result(\ref{1.32}).

{From} the very definition of the integral follows:

\be
\int^x \nabla f= f+c
\label{1.35}
\ee

This can be combined with the Leibniz rule (\ref{1.26})
or (\ref{1.27}) to give a formula for partial integration:

\be
\int^x \nabla fg=fg+c=\int^x (\nabla f)(L^{-1}g) + \int^x (L f) (\nabla g)
\label{1.36}
\ee

or

\be
\int^x \nabla fg=fg+c=\int^x (\nabla f)(Lg) + \int^x (L^{-1} f) (\nabla g)
\label{1.37}
\ee


\section{Field equations in a purely algebraic context}\label{sub2}

Based on the calculus that we have developed in the previous
section we introduce field equations that define the time
development of fields. For this purpose we have to enlarge
the algebra by a central element, the time. Fields now depend
on $x$ and $t$.

We will consider the Schroedinger equation and the Klein-Gordon
equation and demonstrate that there are continuity equations 
for a charge density and for an energy momentum density.

It is also possible to separate space and time dependence to 
obtain the time independent Schroedinger equation. To interpet 
it as an eigenvalue equation a Hilbert space for the solutions
has to be defined. This cannot be done on algebraic grounds only.

\bigskip
\pn
\underline{Schroedinger equation:}

This is the equation of motion that governs the time dependence
of a quantum mechanical system.

To define it we extend the algebra ${\cal{A}}_x$ by the time variable $t$,
in our case it will be a central element and $\overline{t}=t$. We call the
extended algebra ${\cal{A}}_{x,t}$.

The Schroedinger equation acts on a field as follows:

\ba
 i \frac{\partial}{\partial t} \psi(x,t)&=&\Big(-\frac{1}{2m} \nabla^2 +V(x)
\Big) \psi(x,t) \label{2.1} \\
 \psi &\in & {\cal{A}}_{x,t}, \quad V \in {\cal{A}}_{x}, \quad \overline{V}=V
\no
\ea

We show that there is a continuity equation for 

\be
\rho(x,t)=\overline{\psi}(x,t) \psi(x,t)
 \in {\cal{A}}_{x,t}
\label{2.2}
\ee

The continuity equation can be written in the form 

\be
\frac{\partial \rho}{\partial t} + \nabla j=0
\label{2.3}
\ee

This is a consequence of the Schroedinger equation
(\ref{2.1}) and its
conjugate equation. To find the conjugate equation we consider
$\nabla \overline \psi$ and $\overline{\nabla \psi}$ as elements
of ${\cal A}_{x,t}$, where conjugation is defined. We find
$\nabla \overline \psi = \overline {\nabla\psi}$ and therefore

\be
\label{2.4}
-i \frac{\partial}{\partial t} \overline{\psi} = (-\frac{1}{2m}\nabla^2 +V)\overline{\psi}
\ee

This yields

\be
\frac{\partial \rho}{\partial t}=\frac{i}{2m} \left\{ \overline{\psi}
(\nabla^2 \psi)-(\nabla^2 \overline \psi) \psi \right\}
\label{2.5}
\ee

Now we can use Green's theorem (\ref{1.29}) and we obtain:

\be
j=-\frac{i}{2m} L^{-1} \left\{ \overline{\psi}(L \nabla \psi)-
(L \nabla \overline{\psi}) \psi \right\}
\label{2.6}
\ee

The time independent Schroedinger equation can be obtained 
from (\ref{2.1}) by separation:

\be
\psi(x,t)=\varphi(t) U(x)
\label{2.7}
\ee
The usual argument leads to:

\ba
i \frac{\partial}{\partial t} \varphi(t)&=&E \varphi(t), \label{2.8} \\
\left( -\frac{1}{2m} \nabla^2 +V \right) U(x)&=&E U(x), \quad E \in \complex
\no
\ea

This is as far as we can get using the algebra only. To find
meaningful solutions  of (\ref{2.8}) we have to define a linear
space with a norm, a Hilbert space, and the solutions $U(x)$ will have
to be elements of this space.

There is also a continuity equation for the energy momentum density:

\ba
{\cal H}&=&\frac{1}{2mq}(\nabla L^{-1} \overline{\psi})(\nabla L^{-1} \psi) 
+V\overline{\psi}\psi   \label{2.9} \\
\pi &=& \frac{1}{2m}\left( {(\nabla \overline{\psi})(L^{-1} \dot{\psi})+
(L^{-1} \dot{\overline{\psi}}) \nabla \psi}\right)
\label{2.10}
\ea

It follows from (\ref{2.1}) and (\ref{2.3}) that

\be
\label{2.11}
\frac{d}{dt} {\cal H} - \nabla \pi =0
\ee

Both conservation laws will follow from a variational principle that
we shall formulate as soon as we know how to define a definite 
integral. In the meantime it is left as an exercise to prove (\ref{2.9}).

\bigskip
\pn
\underline{Klein-Gordon equation:}

We define

\be
\label{2.12}
\frac{\partial^2}{\partial t^2} \phi - \nabla^2 \phi + m^2 \phi =0
 , \qquad\phi \in {\cal A}_{x,t}
\ee
For complex $\phi$ (\ref{2.12}) and its conjugate lead to current
conservation:

\ba
\label{2.13}
&\rho = \dot{\overline{\phi}} \phi - \overline{\phi}\dot{\phi},
\quad j=-(\nabla \overline{\phi})(L^{-1} \phi) + (L^{-1}\overline{\phi})(\nabla \phi)& \\
&\dot{\rho}+ \nabla j =0&
\label{2.14}
\ea
We have used Green's theorem (\ref{1.29}). The verification of (\ref{2.13})
as well as the verification of energy momentum conservation is again left 
as an exercise.

\ba
{\cal H}&=&\dot{\overline{\phi}}\dot{\phi}+
\frac{1}{q} (\nabla L^{-1} \overline{\phi})
(\nabla L^{-1} \phi) + m^2 \overline{\phi}\phi\label{2.15}\\
\pi &=&(\nabla \overline \phi)(L^{-1}\dot{\phi})+(L^{-1}\dot{\overline{\phi}})(\nabla \phi)\label{2.16}\\[4mm]
&&\dot{\cal H}-\nabla \pi =0\label{2.17}
\ea

\section{Gauge theories in a purely algebraic context}\label{sub3}

It is possible to develop a covariant gauge theory starting from 
fields as defined in the previous chapter. These fields are
supposed to have well-defined  properties under a gauge
transformation. Using the concepts of connection and vielbein
covariant derivatives can be defined. These covariant derivatives form
an algebra and a curvature arises from this algebra in a natural way.

Finally we show how an exterior calculus can be set up that opens the way
to differential geometry on the algebra.

\bigskip
\pn
\underline{Covariant derivatives:}

We assume that $\psi(x,t)$ spans a representation of a compact
gauge group. Let $T_l$ be the generators of the group in this
representation and let $\alpha(x,t)$ be Liealgebra-valued

\ba
\alpha(x,t)=\sum_l g \alpha_l (x,t) T_l \label{3.1} \\
\alpha_l(x,t) \in \A_{x,t}
\no
\ea
We have introduced a coupling constant $g$, $(g \in \complex)$.
 
The field $\psi(x,t)$ is supposed to transform as follows:

\be
\psi'(x,t)=e^{i \alpha(x,t)} \psi(x,t)
\label{3.2}
\ee
A covariant derivative is a derivative such that

\be
(\D \psi)'=e^{i \alpha(x,t)} (\D \psi)
\label{3.3}
\ee
and $\D=\nabla$ for $g=0$.

We make the Ansatz:

\be
\D_x \psi=E (\nabla+ \phi) \psi
\label{3.4}
\ee
Here we have introduced the connection $\phi$ and the 
Vielbein $E$. For $g=0$, $\phi$ has to be zero and $E$ has
to be the unit matrix.

We aim at  a transformation law for $E$ and $\phi$ such that

\be
E'(\nabla+ \phi') e^{i \alpha} \psi=e^{i \alpha} E(\nabla + \phi) \psi 
\label{3.5}
\ee
{From} the Leibniz rule  (\ref{1.26}) for $\nabla$ follows:

\be
\nabla \psi'=(\nabla e^{i \alpha}) L \psi + (L^{-1} e^{i \alpha}) \nabla \psi
\label{3.6}
\ee
Thus (\ref{3.5}) will be satisfied if:

\ba
E'&=&e^{i \alpha} E(L^{-1} e^{-i \alpha}) \label{3.7} \\
\phi'&=&(L^{-1} e^{i \alpha}) \phi (e^{-i \alpha})-
(\nabla e^{i \alpha})(L e^{-i \alpha}) L
\no
\ea

{From} the inhomogeneous terms in the transformation law of
$\phi$ in (\ref{3.7}) we see that $\phi$ has to be $L$-valued.

\be
\phi= g\varphi L
\label{3.8}
\ee
{From} (\ref{3.7}) follows 

\be
g\varphi'=(L^{-1} e^{i \alpha}) g\varphi (L e^{-i \alpha})-(\nabla e^{i \alpha})
(L e^{-i \alpha})
\label{3.9}
\ee

To further analyze (\ref{3.9}) we use the expression (\ref{1.18}) for $\nabla$

\ba
(\nabla e^{i \alpha}) (L e^{-i \alpha})&=&\lambda^{-1} x^{-1}
\Big( (L^{-1}-L) e^{i \alpha} \Big) (L e^{-i \alpha}) \label{3.10} \\
&=& \lambda^{-1} x^{-1} \left[(L^{-1} e^{i \alpha}) (L e^{-i \alpha})-1
\right]
\no
\ea 
This suggests to rewrite (\ref{3.9}) as follows:

\be
g\varphi'- \lambda^{-1} x^{-1}=(L^{-1} e^{i \alpha}) \left[ g\varphi 
-\lambda^{-1} x^{-1} \right] (L e^{-i \alpha})
\label{3.11}
\ee

{From} the transformation law of $E$ (\ref{3.7}) now follows that
the object

\be
E(g\varphi - \lambda^{-1} x^{-1})(LE)= g\chi
\label{3.12}
\ee
transforms homogeneously

\be
\chi'=e^{i \alpha} \chi e^{-i \alpha}
\label{3.13}
\ee
Thus $\chi$ can be chosen to be proportional to
the unit matrix or  to be Liealgebra-valued.

\be
\chi=\chi_01  \quad + \quad \sum_l \chi_l T_l
\label{3.14}
\ee

We are now going to show that $\chi_l=0$ if we demand
a covariant version of (\ref{1.18}).
For this purpose we first have
to find a covariant version of $L$:

\be
(\L \psi)'=e^{i \alpha} \L \psi \quad , \quad \L=L \quad for \quad g=0
\label{3.15}
\ee
We try the Ansatz:

\be
\L=\tilde E L
\label{3.16}
\ee
with a new version of a vielbein $\tilde E$.
{From} (\ref{3.15}) follows

\be
\tilde E'=e^{i \alpha}\tilde E(L e^{-i \alpha}) 
\label{3.17}
\ee

It is natural to identify $\tilde E$ with $(L E^{-1})$ because both have
the same transformation property:

\be
\tilde E=(L E^{-1})
\label{3.18}
\ee

For  $\L$ and its inverse we found:

\be
\L=(L E^{-1})L=L E^{-1}, \quad \L^{-1}=E L^{-1}
\label{3.19}
\ee
$\L^{-1}$ transforms covariant as well.

Now we postulate:

\be
\D_x=\lambda^{-1} x^{-1} (\L^{-1}-\L)
\label{3.20}
\ee
In more detail:

\ba
\D_x &=& \lambda^{-1} x^{-1} \left\{ E (L^{-1}-L) +\left( E-(LE^{-1})\right)
L \right\} \label{3.21} \\
&=& E \nabla + \lambda^{-1} x^{-1} E\left(1-(E^{-1})(LE^{-1}) \right) L
\no
\ea
If we compare this with (\ref{3.4}) we find:

\be
g\varphi=\lambda^{-1} x^{-1} \left(1-E^{-1} (LE^{-1})\right)
\label{3.22}
\ee
The connection is entirely expressed in terms of the vielbein.

Now we show that with this choice of $\varphi$  the covariant
derivative of the vielbein vanishes.

Let us start with a field $H$ that transforms like $E$:

\be
H'=e^{i \alpha} H (L^{-1} e^{-i \alpha})
\label{3.23}
\ee
We apply $\L$ and $\L^{-1}$ to this object:

\ba
\L H=(L E^{-1})(LH) E \label{3.24} \\
\L^{-1} H=E(L^{-1} H) (L^{-1} E^{-1})
\no
\ea
This we insert into (\ref{3.19}) to obtain:

\be
\D_x H=\lambda^{-1} x^{-1} \{ E(L^{-1} H)(L^{-1} E^{-1}) -(LE^{-1})(LH) E \}
\label{3.25}
\ee
If in this formula we substitute $E$ for $H$ we obtain:

\be
\D_x E=0
\label{3.26}
\ee

For the covariant time derivative we follow the standard construction

\be
\D_t=(\partial_t+\omega) \psi, \quad \omega=\sum_l \omega_l(x,t) T_l
\label{3.27}
\ee

The transformation property of $\omega$ is:

\be
\omega'=e^{i \alpha} \omega e^{-i \alpha}+e^{i \alpha} \partial_t e^{-i \alpha}
\label{3.28}
\ee

\vspace{1cm}
\bigskip
\pn
\underline{Curvature:}

The covariant derivatives have an algebraic structure as well.
With the choice (\ref{3.22}) for the connection it follows from
(\ref{3.20}) that

\ba
\L \D_x \psi=q \D_x \L \psi, && \L^{-1} \D_x \psi=q^{-1} \D_x \L^{-1} \psi
\label{3.29} \\[3mm]
(\L \D_t -\D_t \L) \psi &=& \L T \psi \nn
(\L^{-1} \D_t - \D_t \L^{-1}) \psi &=& -T \L^{-1} \psi
\label{3.30}
\ea
$T$ is a tensor quantity:

\be
T=(\partial_t E) E^{-1} -(E)(L^{-1} \omega) E^{-1} + \omega
\label{3.31}
\ee
such that

\be
T'= e^{i \alpha} T e^{-i \alpha}
\label{3.32}
\ee

The commutator of $\D_t$ and $\D_x$ is
easy to compute from (\ref{3.20}) and (\ref{3.29}). We find:

\be
(\D_t \D_x -\D_x \D_t) \psi= T \D_x \psi+ \lambda^{-1} x^{-1}
\{ \L T +T \L^{-1} \} \psi
\label{3.33}
\ee
To avoid the $\lambda^{-1} x^{-1}$ factor we can write this term also
in the form:

\be
\lambda^{-1} x^{-1} \left\{ \L T+ T \L^{-1} \right\}=EF (LE) \L
\label{3.34}
\ee
with 

\be
F=g\partial_t \varphi- \nabla \omega+(L^{-1} \omega)g\varphi-g\varphi (L \omega)
\label{3.35}
\ee
and

\be
E' F' (LE')=e^{i \alpha} EFLEe^{-i \alpha}
\label{3.36}
\ee
The tensor $T$ plays the role of a curvature.

\bigskip
\pn
\underline{Leibniz rule for covariant derivatives:}

We want to learn how  covariant derivatives act on
products of representations. To distinguish the
representations  we are going to use 
indices. Thus $\psi$ and $\chi$ are two representations such that

\be
\psi'_\alpha=(e^{i \alpha})_\alpha{}^{\beta} \psi_\beta, \quad
\chi'_a=(e^{i \alpha})_a{}^b \chi_b
\label{3.37}
\ee
Repeated indices are to be summed. The Vielbein is a matrix
object $E_\alpha{}^\beta$ or $E_a{}^b$, depending on what 
representation it acts on.

\ba
E'_\alpha{}^\beta&=&(e^{i \alpha})_\alpha{}^\sigma E_\sigma {}^\rho
(L^{-1} e^{-i \alpha})_\rho{}^\beta 
\label{3.38} \\
E'_a{}^b&=&(e^{i \alpha})_a{}^s E_s {}^r (L^{-1} e^{-i \alpha})_r{}^b
\no
\ea
On the product of representations it acts as follows:

\be
E_{\alpha a}{}^{\beta b} \psi_\beta \chi_b=E_\alpha{}^\beta E_a{}^b
\psi_\beta \chi_b
\label{3.39}
\ee

This gives us a chance to obtain the vielbein starting from  
fundamental representations.
It is left to  show that such a construction leads to a unique
vielbein for each representation. 

To obtain the Leibniz rule we start with a scalar:

\be
f'=f
\label{3.40}
\ee
In this case the  derivatives are the covariant derivatives:

\be
\D_x f=\nabla f, \quad \D_t f=\frac{\partial}{\partial t} f, \quad
\L f=L f, \quad \L^{-1} f=L^{-1} f
\label{3.41}
\ee
If we combine this representation with $\psi$ we obtain

\ba
(\L f \psi)_{\alpha}&=&(L E^{-1})_\alpha{}^\beta(L f \psi)_\beta=(Lf)
(L E^{-1})_\alpha{}^\beta (L \psi)_\beta \label{3.42} \\
&=&\L f(\L \psi)_\alpha
\no
\ea

As a second example we treat the scalar obtained from a covariant
and a contravariant representation.

\be
\psi'=e^{i\alpha}\psi \quad , \quad \chi'=\chi -e^{i\alpha}
\label{3.43}
\ee
such that 

\be
\chi'\psi'=
\chi \psi
\label{3.44}
\ee
We take the covariant action of $L$ on  $\chi \psi$:

\ba
\L (\chi^\alpha \psi_\alpha) 
&=& L (\chi^\alpha \psi_\alpha)
=(L \chi^\alpha) (L \psi_\alpha) \nn
&=&(L \chi^\sigma)(LE)_\sigma{}^\alpha(L E^{-1})_\alpha{}^\beta
(L \psi)_\beta \label{3.45} \\
&=& (\L \chi)^\alpha (\L \psi)_\alpha \no
\ea

Encouraged by this result we continue with the product of two 
arbitrary representations:

\ba
\L (\psi \chi)_{\alpha a}&=&(L E^{-1})_\alpha{}^\beta{}_a{}^b
(L \psi_\beta) (L \chi_b) \nn
&=& (L E^{-1})_\alpha {}^\beta (L E^{-1})_a {}^b (L \psi)_\beta (L \chi)_b 
\label{3.46} \\
&=& (\L \psi)_\alpha (\L \chi)_a \no
\ea

This is the Leibniz rule for the covariant version
of $L$. It is obvious that the same holds for $\L^{-1}$:

\be
\L^{-1} \psi_\alpha \chi_a =(\L^{-1} \psi)_\alpha(\L^{-1} \chi)_a
\label{3.47}
\ee

The covariant derivative  $\D_x$ can be obtained  from $\L$
and $\L^{-1}$  according to the same argument that led to
the Leibniz rule for the derivative $\nabla$ (\ref{1.26}),
we can now show that:

\be
\D_x (\psi \chi) =(\D_x \psi) (\L \chi)+(\L^{-1} \psi) (\D_x \chi)
\label{3.48} 
\ee
or 

\be
\D_x (\psi \chi)=(\D_x \psi) (\L^{-1} \chi)+(\L \psi) (\D_x \chi)
\label{3.49}
\ee
This is the Leibniz rule for covariant derivatives.

The Leibniz rule (\ref{3.47}), (\ref{3.48}) and 
(\ref{3.49}) allow us to drop the field $\psi$ in the
equations (\ref{3.29}) and (\ref{3.34}) and write them as 
algebra relations.

\ba
\L \D_x=q \D_x \L, \quad \L^{-1} \D_x=q^{-1} \D_x \L^{-1} \nn
\L \D_t -\D_t \L=\L T \label{3.50} \\
\L^{-1} \D_t -\D_t \L^{-1}=-T \L^{-1} \nn
\D_t \D_x-\D_x \D_t=T \D_x+EF(LE)\L \no
\ea

$T$ plays the role of an independent tensor that is a function 
of the vielbein and the connection $\omega$. It is defined in 
equation (\ref{3.31}). $F$ depends on $T$ and is defined in 
equation (\ref{3.34}).

In a later chapter we shall see that the vielbein can be expressed
in terms of a connection $A_{\mu}$ in the usual definition of
a covariant derivative.

\bigskip
\pn
\underline{Exterior Derivative:}

To define differentials we have to extend the algebra by an element 
$dx$. We assume for the moment that there is an algebraic relation
that allows us to order $x$ and $dx$:

\be
dx f(x)=a[f(x)] dx, \quad a: {\cal A}_x \rightarrow {\cal A}_x
\label{3.51}
\ee

The map $a$ has to be an automorphism of the algebra $\A_x$

\be
a[f(x)g(x)]=a[f(x)]a[g(x)]
\label{3.52}
\ee
This can be seen as follows

\ba
dx \; f \times g&=& a[f \times g] dx \label{3.53} \\
&=& a[f] dx g=a[f] a[g] dx
\label{3.54}
\ea

The exterior derivative $d$ can be defined as follows:

\be
d \; x= dx, \quad d=dx \nabla, \quad d^2=0
\label{3.55}
\ee

{From} (\ref{1.11}) follows:

\be
d \: x^m=[m] dx x^{m-1}
\label{3.56}
\ee
This tells us how $d$ acts on any field $f(x) \in {\cal{A}}_x$.

To derive a Leibniz rule for the exterior derivative we use the
Leibniz rule for the derivative (\ref{1.26})

\be
d \; fg=dx  \{(\nabla f) (L^{-1} g)+(L f) \nabla g \}
\label{3.57}
\ee
The relation (\ref{1.36}) allows us to obtain a Leibniz rule for $d$:

\be
d \; fg = (df)(L^{-1} g)+a[Lf] dg
\label{3.58}
\ee
The simplest choice is:

\be
a[f]=f, \quad dx \; x =x \; dx
\label{3.59}
\ee
Then we obtain from (\ref{3.56})

\be
d \; fg=(df)(L^{-1}g)+(Lf) dg
\label{3.60}
\ee

We could have used the Leibniz rule 
(\ref{3.58}), then we would have obtained:

\be
d \; fg =dx \{ (\nabla f)(Lg)+(L^{-1} f) \nabla g \}
\label{3.61}
\ee
For $a(f)$ defined by (\ref{3.59}) this yields:

\be
d \; fg =(df)(Lg)+(L^{-1} f) dg
\label{3.62}
\ee

A different choice for $a[f]$ could be 

\be
a[f]=L^p f, \quad p \in \zet
\label{3.63}
\ee
This satisfies (\ref{3.52}).

The Leibniz rule (\ref{3.58}) then yields

\be
d \; fg=(df)(L^{-1} g)+(L^{p+1} f) (dg)
\label{3.64}
\ee
or

\be
d \; fg=(df)(L g)+(L^{p-1} f) (dg)
\label{3.65}
\ee

\section{$q$-Fourier transformations}\label{sub4}

In the next chapter we will study representations of the algebra 
(\ref{1.1}). As mentioned before we are interested in ``good''
representations where the coordinates and the momenta can be 
diagonalized. We also want to have explicit formulas for the 
change of the coordinate bases to the momentum bases. This is
the $q$-Fourier transformation and it turns out that the
relevant transition functions are the well known 
$q$-functions $cos_qx$ and $sin_qx$. We are going to 
discuss these functions now. It should be mentioned that 
in addition to quantum groups there is another reservoir of detailed
mathematical knowledge - these are the basic hypergeometric series,
special example: $sin_q(x)$ and $cos_q(x)$.

The $q$-Fourier transformation is based on the $q$-deformed 
$cosine$ and $sine$ functions that are defined as follows:
\ba
\cos_q(x)&=&\sum^{\infty}_{k=0}(-1)^k\frac{x^{2k}}{[2k]!}\frac{q^{-k}}{\lambda^{2k}}\label{4.1}\\
\sin_q(x)&=&\sum^{\infty}_{k=0}(-1)^k\frac{x^{2k+1}}{[2k+1]!}\frac{q^{k+1}}{\lambda^{2k+1}}\no
\ea

the symbol $[n]$ was defined in (\ref{1.11}), and $[n]!$  stands for: 

\be
\label{4.2}
[n]!=[n]\,[n-1]\;\cdots\;[1]
\ee
\ba
[n]=\frac{q^n-q^{-n}}{q-\frac{1}{q}}=q^{n-1}+q^{n-3}+\;\ldots\;+q^{-n+3}+q^{-n+1}
\no
\ea

These $cos_q$ and $sin_q$ functions are solutions of the equations:

\ba
\frac{1}{x}\left(\sin_q(x)-\sin_q(q^{-2}x)\right)&=&\cos_q(x)\label{4.3}\\
\frac{1}{x}\left(\cos_q(x)-\cos_q(q^{-2}x)\right)&=&-q^{-2}\sin_q(q^{-2}x)\no
\ea

The $cos_q(x)$ and $sin_q(x)$ functions (\ref{4.1}) are determined by these
equations up to an overall normalization. To prove this is
straightforward, as an example we verify the first of the equations
(\ref{4.3})

\be
\label{4.4}
\frac{1}{x}\left(\sin_q(x)-\sin_q(q^{-2}x)\right)=\sum^{\infty}_{k=0}(-1)^k\frac{x^{2kx}}{[2k+1]!}\frac{q^{k+1}}{\lambda^{2k+1}}(1-q^{-2(2k+1)})
\ee

but

\be
\label{4.5}
1-q^{-2(2k+1)}=\frac{\lambda}{q}\:q^{-2k}[2k+1]
\ee

and (\ref{4.3}) follows. The relations (\ref{4.3}) are the analogon 
to the property of the usual $cos$ and $sin$ functions that the
derivative of $sin (cos)$ is $cos (-sin)$.

The most important property of the $cos_q$ and $sin_q$ functions is that they
each form a complete and orthogonal set of functions. This allows us
to formulate the $q$-Fourier theorem for functions $g(q^{2n})$ that are
defined on the $q$-lattice points $q^{2n},\quad n\in\zet$.
\ba
\tilde{g}_c(q^{2\nu})&=&N_q\sum_{n=-\infty}^{\infty}q^{2n}\cos_q(q^{2(\nu+n)})
\,g(q^{2n})\label{4.6}\\
g(q^{2n})&=&N_q\sum_{\nu=-\infty}^{\infty}q^{2\nu}\cos_q(q^{2(\nu+n)})
\,\tilde{g}_c(q^{2\nu})\no
\ea

and:

\be
\sum_{n=-\infty}^{\infty}q^{2n}\left|g(q^{2n})\right|^2=\sum_{\nu=-\infty}^{\infty}q^{2\nu}\left|\tilde{g}_c(q^{2\nu})\right|^2\label{4.7}
\ee

The normalization constant $N_q$ can be calculated:

\be
N_q=\prod_{\nu=0}^{\infty}\left(\frac{1-q^{-2(2\nu+1)}}{1-q^{-4(\nu+1)}}\right),\quad q>1
\label{4.8}
\ee

Another transformation is obtained by using $sin_q$ instead of $cos_q$:

\ba
\tilde{g}_s(q^{2\nu})&=&N_q\sum_{n=-\infty}^{\infty}q^{2n}\sin_q(q^{2(\nu+n)})
\,g(q^{2n})\label{4.9}\\
g(q^{2n})&=&N_q\sum_{\nu=-\infty}^{\infty}q^{2\nu}\sin_q(q^{2(\nu+n)})
\,\tilde{g}_s(q^{2\nu})
\no
\ea

and

\be
\sum_{n=-\infty}^{\infty}q^{2n}\left|g(q^{2n})\right|^2=\sum_{\nu=-\infty}^{\infty}q^{2\nu}\left|\tilde{g}_s(q^{2\nu})\right|^2\label{4.10}
\ee

That $sin_q$ and $cos_q$ individually form a complete set of functions can
be understood because the range of $x=q^{2n}$ is restricted to $x\ge 0$.

The orthogonality and completeness property can be stated as 
follows:

\ba
N_q^2\sum_{\nu=-\infty}^{\infty}
q^{2\nu}\cos_q(q^{2(n+\nu)})\cos_q(q^{2(m+\nu)})&=&q^{-2n}\,\delta_{n,m}\label{4.11}\\
N_q^2\sum_{\nu=-\infty}^{\infty}
q^{2\nu}\sin_q(q^{2(n+\nu)})\sin_q(q^{2(m+\nu)})&=&q^{-2n}\,\delta_{n,m}
\no
\ea

That the role of $n$ and $\nu$ can be exchanged to get the 
completeness (orthogonality) relations from the orthogonality
(completeness) relations is obvious.

There is also a deformation of the relation $cos^2+sin^2=1$.
It is:

\be
\cos_q(x)\cos_q(qx)+q^{-1}\sin_q(x)\sin_q(q^{-1}x)=1
\label{4.12}
\ee

the general term in the $cos_q$ product is:

\be
\cos_q(x)\cos_q(qx)=\sum_{l,k=0}^{\infty}
\left(\frac{x}{\lambda}\right)^{2(k+l)}(-1)^{k+l}\frac{q^{l-k}}{[2k]!\,[2l]!}
\label{4.13}
\ee

and in the $sin_q$ product it is:

\be
q^{-1}\sin_q(x)\sin_q(q^{-1}x)=\sum_{l,k=0}^{\infty}
\left(\frac{x}{\lambda}\right)^{2(k+l+1)}(-1)^{k+l}\frac{q^{l-k}}{[2k+1]!\,[2l+1]!}
\label{4.14}
\ee

The terms of  (\ref{4.12}) that add up to a fixed power $2n$ of
$x$ are

\be
(-1)^n\left(\frac{x}{\lambda}\right)^{2n}\frac{1}{[2n]!}\sum_{k=0}^{2n}
(-1)^k q^{n-k}\frac{[2n]!}{[k]![2n-k]!}\label{4.15}
\ee

For $n=0$ this is one, for $n=1$ it is:

\be
-\left(\frac{x}{\lambda}\right)^2 \frac{1}{[2]}\left\{q-[2]+\frac{1}{q}\right\}
=0
\label{4.16}
\ee

For the general term (\ref{4.12}) was proved by J.Schwenk.

In the formulas for the Fourier transformations  (\ref{4.6}) and
(\ref{4.9}) only even powers of $q$ enter whereas (\ref{4.12})
connects even powers to odd powers. From (\ref{4.11}) we see that
$cos_q(q^{2n})$ and $sin_q(q^{2n})$ have to tend to zero for $n\rightarrow \infty$. For (\ref{4.12})
to hold $cos_q(q^{2n+1})$ and $sin_q(q^{2n+1})$ have to diverge for $n\rightarrow \infty$. This behaviour is
illustrated by Fig.1. and Fig.2. They show that $cos_q(x)$ and $sin_q(x)$ are 
functions that diverge for $x\rightarrow \infty$, they are not functions of $L^2$. 
However, the points $x=q^{2n}$ are close to the zeros of $cos_q(x)$ and $sin_q(x)$
and for $n\rightarrow \infty$ tend to these zeros such that the sum in
(\ref{4.11}) is convergent.

We can also consider $cos_q(x)$ and $sin_q(x)$ as a field, i.e. as an
element of the algebra ${\cal A}_x$. Then we can apply $\nabla$ to it.
We use (\ref{1.18}) as an expression for $\nabla$:

\be
\nabla cos_q(kx)=\frac{1}{\lambda}\,\frac{1}{x}
\left\{\cos_q(qkx)-\cos_q(q^{-1}kx)\right\}\label{4.17}
\ee

Now we use (\ref{4.3}) for the variable $y=qkx$ and we obtain

\be
\nabla cos_q(kx)=-k\,\frac{1}{q\lambda}\sin_q(q^{-1}kx)
\label{4.18}
\ee

The same can be done for $sin_q(kx)$:

\be
\nabla sin_q(kx)=k\,\frac{q}{\lambda}\cos_q(qkx)
\label{4.19}
\ee

This  shows that $cos_q(kx)$ and $sin_q(kx)$ are eigenfunctions of $\nabla^2$:

\ba
\nabla^2\cos_q(kx)&=&-\frac{k^2}{q\lambda^2}\cos_q(kx)\label{4.20}\\
\nabla^2\sin_q(kx)&=&-\frac{k^2 q}{\lambda^2}\sin_q(kx)\no
\ea

We have found eigenfunctions for the free Schroedinger equation
(\ref{2.8}) $(V=0)$. To really give a meaning to them we
have to know how to define a Hilbert space for the solutions.

\begin{figure}
\centerline{\epsfig{figure=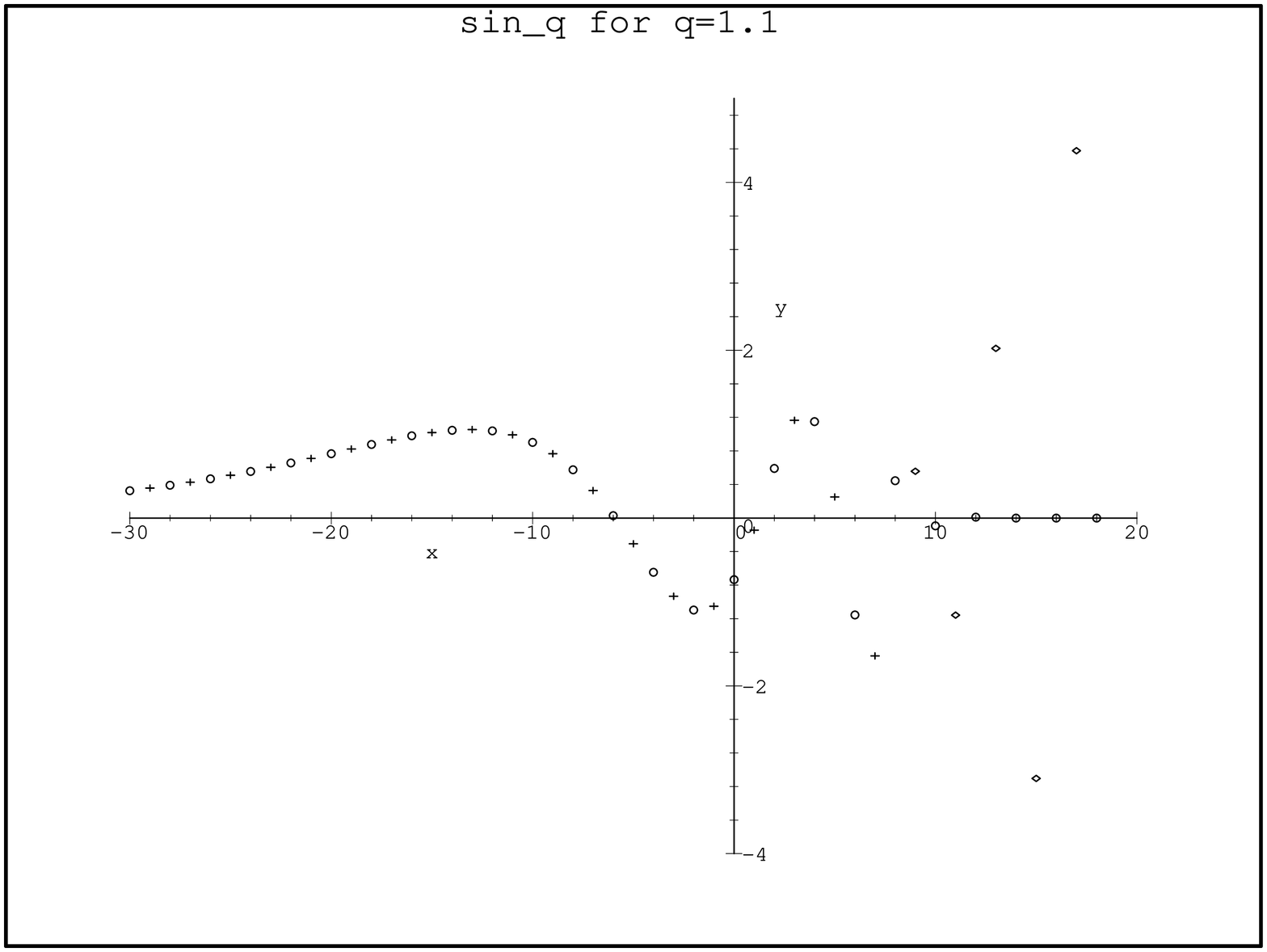,height=130mm,angle=270}}
\caption{$sin_q(q^{n})$ for $q=1.1$.
Crosses (circles) indicate odd (even) $n$.
For $n>8$ a logarithmic $y$-scale was used.}
\end{figure} 

\begin{figure}
\centerline{\epsfig{figure=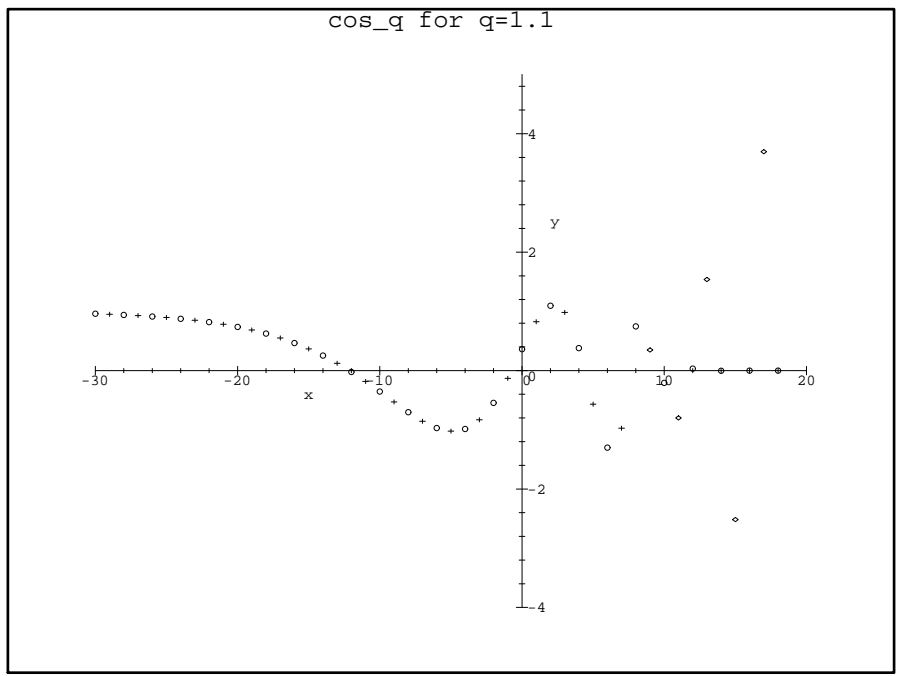,height=130mm,angle=270}}
\caption{$cos_q(q^{n})$ for $q=1.1$.
Crosses (circles) indicate odd (even) $n$.
For $n>8$ a logarithmic $y$-scale was used.}
\end{figure}

\section{Representations}\label{sub5}

In the first three chapters we have developed a formalism that is
entirely based on the algebra. For a physical interpretation
we have to relate this formalism to real numbers - real numbers
being the result of measurements. This can be done by studying
representations of the algebra and adopting the interpretation
scheme of quantum mechanics. Thus we have to aim at representations
where selfadjoint elements of the algebra that correspond to physical
observables like position or momentum are represented by 
(essentially-) 
selfadjoint linear operators in a Hilbert space. We want
to diagonalize these operators, they should have real eigenvalues
and the eigenvectors should form a basis in a Hilbert space. 
We shall call these representations ``good'' representations.

\bigskip
\pn
\underline{The Hilbert space ${\cal H}_s^{\sigma} $:}

We first represent the algebra
\begin{equation}
 \label{5.1}
 x\Lambda=q\Lambda x,\quad \bar{x}=x,\quad \bar{\Lambda}=\Lambda^{-1}
\end{equation} 
From what was said above we can assume $x$ to be diagonal.
From (\ref{5.1}) follows that with any eigenvalue $s$ of the 
linear operator $x$ there will be the eigenvalues  $q^ns,\quad n\in {\bf Z}$.
As $s$ is the eigenvalue of a selfadjoint operator it is real.
We shall restrict $s$ to be
positive and use $-s$ for negative eigenvalues. In the
following we will assume $q>1$. 

We start with the
following eigenvectors and eigenvalues:

\begin{eqnarray}
\label{5.2}
x|n,\sigma\rangle^s=\sigma sq^n      |n,\sigma\rangle^s \\
n\in \zet,\quad \sigma =\pm 1,\quad 1\le s<q\nonumber
\end{eqnarray}
The algebra (\ref{5.1}) is represented on the states with 
$\sigma$ and $s$ fixed:

\begin{equation}
\label{5.3}
\Lambda |n,\sigma \rangle^s=|n+1,\sigma \rangle^s
\end{equation}

A possible phase is absorbed in the definition of the states.
These states are supposed to form an orthonormal basis in a
Hilbert space, which we will call ${\cal H}_s^{\sigma}$.

\begin{eqnarray}
\label{5.4}
|n,\sigma \rangle^s \in {\cal H}_s^{\sigma} \\
^s\langle m,\sigma|n,\sigma \rangle^s=\delta_{n,m}
\no
\end{eqnarray}
This makes $\Lambda$, defined by (\ref{5.3}), a unitary linear operator.

To obtain a representation of the algebra (\ref{1.1}) we use
formula (\ref{1.7}) to represent $p$. We find an irreducible 
representation of the $x, \Lambda, p$ algebra.

\be
p\ket{n,\sigma}^s=i\lambda^{-1}\frac{\sigma}{s}q^{-n}\left\{ q^{-\frac{1}{2}}\ket{n+1,\sigma}^s-q^{\frac{1}{2}}\ket{n-1,\sigma}^s\right\}
\label{5.5}
\ee
This defines the action of $p$ on the states of the bases (\ref{5.4}).
A direct calculation shows that the linear operator $p$ defined by
(\ref{5.5}) indeed satisfies the algebra (\ref{1.1}) and that $p$
is hermitean.

\be
\overline{\bra{n+1}p\ket{n}}=\bra{n}p\ket{n+1}=-i\lambda^{-1}q^{-n-\frac{1}{2}}
\label{5.6}
\ee

However, $p$ is not selfadjoint. 
There are no selfadjoint extensions of $p$ in the 
Hilbertspace defined by (\ref{5.4}) with the sign $\sigma$ fixed such
that $\Lambda p \Lambda^{-1}=qp$.
To show this we assume $p$ to be 
selfadjoint, i.e. diagonizable with real eigenvalues  and eigenstates
that form a basis of the Hilbert space (\ref{5.4}). At the same time
the algebra (\ref{1.1}) should be represented.

\be
p\ket{p_0}=p_0\ket{p_0},\;\ket{p_0}=\sum_n c_n^{p_0}\ket{n,\sigma}^s
\label{5.7}
\ee

From the algebra follows that $\Lambda\ket{p_0}$ and $\Lambda^{-1}\ket{p_0}$ are orthogonal to $\ket{p_0}$
because they belong to different eigenvalues of $p$.

\ba
p\Lambda \ket{p_0}&=&\frac{1}{q}p_0\Lambda\ket{p_0}\label{5.8}\\
p\Lambda^{-1} \ket{p_0}&=&qp_0\Lambda^{-1}\ket{p_0}\no
\ea
We conclude

\be
\bra{p_0}\Lambda\ket{p_0}=\bra{p_0}\Lambda^{-1}\ket{p_0}=0
\label{5.9}
\ee
for every eigenvalue $p_0$ of $p$.

From the algebra (\ref{1.1}) follows:

\be
\bra{p_0}\left(q^{\frac{1}{2}}xp-q^{-\frac{1}{2}}px\right)\ket{p_0}=p_0(q^{\frac{1}{2}}-q^{-\frac{1}{2}})\bra{p_0}x\ket{p_0}=0
\label{5.10}
\ee
If there is a $p_0\neq 0$ we find

\be
\bra{p_0}x\ket{p_0}=0
\label{5.11}
\ee
Now we use the representation of $\ket{p_0}$ in the $\ket{n,\sigma}^s$ basis 
to conclude

\be
\sigma \sum_n q^n\left|c_n^{p_0}\right|^2=0
\label{5.12}
\ee
For fixed $\sigma$ this can only be true if all the
$c_n^{p_0}$ vanish. A clear contradiction.

We have shown that $p$ as defined by (\ref{5.5}) is not a 
selfadjoint linear operator. It does, however, satisfy the algebra (\ref{1.1}).
We are now going to show that it has selfadjoint extensions. I would like to 
thank Professor Schmüdgen for pointing out to me that $p$ is
given by a Jacobi matrix and thus has a selfadjoint extension.
We show that there is a basis in ${\cal H}_s^{\sigma}$ where $p$ is diagonal and has
real eigenvalues. We define the states:

\ba
\ket{\tau p_{\nu},\sigma}_I^s&=&\frac{1}{\sqrt{2}}N_q\sum_{n=-\infty}^{\infty}q^{n+\nu}\left\{\cos_q(q^{2(n+\nu)})\ket{2n,\sigma}^s\right.\label{5.13}\\
&&\hspace{2cm}\left.+\tau i\sin_q(q^{2(n+\nu)})\ket{2n+1,\sigma}^s \right\}\no\\
\tau &=&\pm 1\no
\ea

First we prove orthogonality with the help of (\ref{4.11}):

\ba
{}_I^s\bra{\tau' p_{\mu},\sigma}&&\hspace{-5mm}\tau p_\nu ,\sigma\rangle_I^s\no\\
&=&\frac{1}{2}N_q^2\sum_{n=-\infty}^{\infty}q^{2n+\nu+\mu}\left(\cos_q(q^{2(n+\nu)})\cos_q(q^{2(n+\mu)})\right.\no\\
&&\hspace{2cm}\left.+\tau \tau'\sin_q(q^{2(n+\nu)})\sin_q(q^{2(n+\mu)})\right)\no\\
&=&\frac{1}{2}\delta_{\nu\mu}(1+\tau \tau')=\delta_{\nu\mu}\delta_{\tau{\tau'}}
\label{5.14}
\ea

Next we prove completeness. First we form a linear combination of the states (\ref{5.13}) to be able to use (\ref{4.11}) again:

\be
N_q\sum_n q^{n+\nu}\cos_q(q^{2(n+\nu)})\ket{2n,\sigma}^s=\frac{1}{\sqrt{2}}\left\{\ket{p_{\nu},\sigma}_I^s+\ket{-p_{\nu},\sigma}_I^s\right\}
\label{5.15}
\ee

and find

\be
N_q\sum_\nu q^{m+\nu}\cos_q(q^{2(m+\nu)})\frac{1}{\sqrt{2}}\left\{\ket{p_{\nu},\sigma}_I^s+\ket{-p_{\nu},\sigma}_I^s\right\}=
\ket{2m,\sigma}^s
\label{5.16}
\ee

and similarly:

\be
N_q\sum_{\nu} q^{m+\nu}\cos_q(q^{2(m+\nu)})\frac{(-i)}{\sqrt{2}}\left\{\ket{p_{\nu},\sigma}_I^s-\ket{-p_{\nu},\sigma}_I^s\right\}=
\ket{2m+1,\sigma}^s
\label{5.17}
\ee

Thus the states of (\ref{5.13}) form a basis. We finally 
show that they are eigenstates of $p$. We have to use
(\ref{4.3}) and we obtain

\be
p\ket{\tau p_\nu ,\sigma}_I^s=\frac{1}{s\lambda q^{\frac{1}{2}}}\sigma\tau q^{2\nu}\ket{\tau p_\nu ,\sigma}_I^s
\label{5.18}
\ee
these are eigenstates of $p$ with positive and negative
eigenvalues but with eigenvalues spaced by $q^2$. From
the algebra that includes $\Lambda$ 
as well we expect a spacing by $q$ as was shown in 
(\ref{5.8}). We conclude that the selfadjoint extension of
$p$ does not represent the algebra  (\ref{1.1}).
This is in agreement with our previous arguments.

We can apply $\Lambda$ to the state (\ref{5.13}) and we obtain:

\ba
\Lambda\ket{\tau p_\nu ,\sigma}_I^s&=&\frac{1}{\sqrt{2}}N_q\sum_{n=-\infty}^{\infty}q^{n+\nu}\left\{\cos_q(q^{2(n+\nu)})\ket{2n+1,\sigma}^s\right.\no\\
&&\hspace{2cm}\left. +i\tau q^{-1}\sin_q(q^{2(n+\nu-1)})\ket{2n,\sigma}^s \right\}\no\\
&=&\ket{\tau p_\nu ,\sigma}_{II}^s
\label{5.19}
\ea

This defines the states $\ket{\tau p_\nu ,\sigma}_{II}^s$.
They again form a basis and are eigenstates of $p$
\be
p\ket{\tau p_\nu ,\sigma}_{II}^s=\frac{1}{s\lambda q^{\frac{1}{2}}}\sigma \tau q^{2\nu-1}\ket{\tau p_\nu ,\sigma}_{II}^s
\label{5.20}
\ee
This can be shown in the same way as for the states 
(\ref{5.13}). These states define a different selfadjoint 
extension of $p$. Now the eigenvalues differ by a factor $q$ 
from the previous eigenvalues. 

We note that the sign of the eigenvalues depends on $\sigma \tau$
and it is positive and negative for $\sigma$ positive as well 
as for $\sigma$ negative. This suggests to start from 
reducible representations ${\cal H}_s^+\oplus{\cal H}_s^-$
and to extend $p$ in the reducible representations. We 
define

\ba
\ket{\tau p_\nu}_{I}^s&=&\frac{1}{\sqrt{2}}\left\{\ket{\tau p_\nu,+}_{I}^s+\ket{-\tau p_\nu,-}_{I}^s\right\}\label{5.21}\\
&=&\frac{1}{2}N_q\sum_{n=-\infty}^{\infty}q^{n+\nu}\left\{\cos_q(q^{2(n+\nu)})(\ket{2n,+}^s+\ket{2n,-}^s)\right.\no\\
&&\hspace{2cm}\left. +i\tau\sin_q(q^{2(n+\nu)})(\ket{2n+1,+}^s-\ket{2n+1,-}^s) \right\}
\no
\ea

These are eigenstates of $p$:

\be
p\ket{\tau p_\nu}_{I}^s=\frac{1}{s\lambda q^{\frac{1}{2}}}\tau q^{2\nu}\ket{\tau p_\nu}_{I}^s
\label{5.22}
\ee

They are orthogonal because the $\sigma =+1$ states
are orthogonal to the $\sigma =-1$ states.
By the same analysis as for (\ref{5.16}), (\ref{5.17}) we obtain
the states 

\ba
&&\frac{1}{\sqrt{2}}\left\{\ket{2m,+}^s+\ket{2m,-}^s)\right\}\no\\
\mbox{and}\hspace{1cm}&&\frac{1}{\sqrt{2}}\left\{\ket{2m+1,+}^s-\ket{2m+1,-}^s)\right\}
\label{5.23}
\ea
To obtain all the states in ${\cal H}_s^+\oplus{\cal H}_s^-$ we add the states

\ba
\ket{\tau p_\nu}_{II}^s&=&\frac{1}{\sqrt{2}}\left\{\ket{\tau p_\nu,+}_{II}^s+\ket{-\tau p_\nu,-}_{II}^s\right\}\label{5.24}\\
&=&\frac{1}{2}N_q\sum_{n=-\infty}^{\infty}q^{n+\nu}\left\{\cos_q(q^{2(n+\nu)})(\ket{2n+1,+}^s+\ket{2n+1,-}^s)\right.\no\\
&&\hspace{2cm}\left. +i\tau q^{-1}\sin_q(q^{2(n+\nu -1)})(\ket{2n,+}^s-\ket{2n,-}^s) \right\}\no
\ea

They are eigenstates of $p$:

\be
p\ket{\tau p_\nu}_{II}^s=\frac{1}{s\lambda q^{\frac{1}{2}}}\tau q^{2\nu-1}\ket{\tau p_\nu}_{II}^s
\label{5.25}
\ee

They allow us to obtain the states 

\ba
&&\frac{1}{\sqrt{2}}\left(\ket{2m,+}^s-\ket{2m,-}^s)\right)\no\\
\mbox{and}\hspace{1cm}  
&&\frac{1}{\sqrt{2}}(\ket{2m+1,+}^s+\ket{2m+1,-}^s))
\ea
by a Fourier transformation. 

The states defined in (\ref{5.21}) and (\ref{5.24}) form a complete set of
states. The states defined in (\ref{5.21}) are orthogonal
by themselves as well as the states defined in (\ref{5.24}). It 
remains to be shown that the states defined in (\ref{5.21}) are
orthogonal to the states defined in (\ref{5.24}). But this is
obvious because

\be
\left\{ {}^s\bra{n,+}+{}^s\bra{n,-}\right\}\left\{\ket{m,+}^s-\ket{m,-}^s\right\}=0
\label{5.26}
\ee

The states $\ket{\tau p_\nu}_{I}^s$ and $\ket{\tau p_\nu}_{II}^s$
form a basis in ${\cal H}_s^+\oplus{\cal H}_s^-$, they 
are eigenstates of $p$:
\ba
p\ket{\tau p_\nu}_{I}^s&=&\frac{1}{s\lambda q^{\frac{1}{2}}}q^{2\nu}\ket{\tau p_\nu}_{I}^s\no\\
p\ket{\tau p_\nu}_{II}^s&=&\frac{1}{s\lambda q^{\frac{1}{2}}}q^{2\nu-1}\ket{\tau p_\nu}_{II}^s\label{5.27}
\ea

and $\Lambda$ is defined on them and it is unitary.

\ba
\Lambda\ket{\tau p_\nu}_{I}^s&=&\ket{\tau p_\nu}_{II}^s\no\\
\Lambda\ket{\tau p_\nu}_{II}^s&=&\ket{\tau p_{\nu-1}}_{I}^s\label{5.28} 
\ea

This representation of the algebra in terms of selfadjoint linear
operators $x$ and $p$ is irreducible despite the fact that it has 
been built on the direct sum of the two Hilbertspaces 
${\cal H}_s^+\oplus{\cal H}_s^-$.sshssh

A dynamics can now be formulated by defining a Hamilton
operator in terms of $p,x$ and $\Lambda$. The simplest Hamiltonian is

\be
H=\frac{1}{2} p^2
\label{5.29}
\ee

We have calculated its eigenvalues and its eigenfunctions. From the
analysis of the $cos_q(x)$ and $sin_q(x)$ functions and from Fig.1. and Fig.2. we know
that these states get more squeezed with increasing energy.  

\newpage
\section{The definite integral and the Hilbert space $L^2{}_q$}\label{sub6}

\bigskip
\pn
\underline{The definite integral}

Within a given representation of the algebra (\ref{1.1})
a definite integral can be defined. We consider the 
representation where $s$ of eqn (\ref{5.2}) is one
$s=1$ and $\sigma=+1$. The states we shall denote by
$\ket{n,+}$.

We define the definite integral as the difference of matrix
elements of the indefinite integral (\ref{1.30}):

\be
\int_N^Mf(x)=\bra{+,M}\int^x\!\!f(x)\:\ket{M,+}-\bra{+,N}\int^x\!\!f(x)\:\ket{N,+}
\label{6.1}
\ee

From (\ref{1.35}) follows:

\be
\int_N^M\nabla f=\bra{+,M}\:f\:\ket{M,+}-\bra{+,N}\:f\:\ket{N,+}
\label{6.2}
\ee

We can use (\ref{6.1}) for monomials to integrate power
series in $x$:

\be
\int_N^M x^n=\frac{1}{[n+1]}\left(q^{M(n+1)}-q^{N(n+1)}\right)
\label{6.3}
\ee
or we can use (\ref{1.33}) where we defined $\nabla^{-1}$
and take the appropriate matrix elements

\ba
\bra{+,M}\int^x\!\!f(x)\:\ket{M,+}&=&\bra{+,M}\:\nabla^{-1}f(x)\:\ket{M,+}\no\\
&=&\lambda\sum_{\nu=0}^\infty\bra{+,M}\,L^{2\nu}Lx\,f(x)\ket{M,+}
\label{6.4}
\ea
The action of $L$ can be replaced by the operator $\Lambda$:

\be
\bra{+,M}\int^x\!\!f\:\ket{M,+}=\lambda\sum_{\nu=0}^\infty\bra{+,M}\Lambda^{2\nu}Lxf(x)\Lambda^{-2\nu}\ket{M,+}
\label{6.5}
\ee
and $\Lambda$ can now act on the states:
\be
\bra{+,M}\int^xf\ket{M,+}=\lambda\sum_{\nu=0}^\infty\bra{+,M-2\nu}Lxf(x)\ket{M-2\nu,+}
\label{6.6}
\ee

This suggests treating the matrix elements for even and
odd values of $\mu$ separately:
\ba
\bra{+,2M}\int^xf\ket{2M,+}&=&\lambda\sum_{\mu=-\infty}^M\bra{+,2\mu}Lxf(x)\ket{2\mu,+}
\label{6.7}\\
\bra{+,2M+1}\int^xf\ket{2M+1,+}&=&\lambda\sum_{\mu=-\infty}^M\bra{+,2\mu+1}Lxf(x)\ket{2\mu+1,+}
\no
\ea

The definite integral (\ref{6.1}) from even (odd) $N$ to even
(odd) $M$ now finds its natural form:

\ba
\int_{2N}^{2M}\!f(x)&=&\lambda\sum_{\mu=N+1}^M\bra{+,2\mu}Lxf(x)\ket{2\mu,+}\label{6.8}\\
\int_{2N+1}^{2M+1}\!f(x)&=&\lambda\sum_{\mu=N+1}^M\bra{+,2\mu+1}Lxf(x)\ket{2\mu+1,+}\
\no
\ea
These are Riemannian sums for the Riemannian integral.

With the formula (\ref{6.7}) it is possible to integrate
$x^{-1}$ as well:

\be
\int_{2N}^{2M}\frac{1}{x}=\lambda(M-N)
\label{6.9}
\ee
If we take the limit $q\rightarrow 1$ and define $\overline x=q^{2M}$ and $\underline x=q^{2N}$
(\ref{6.9})  approaches the formula

\be
\int_{\underline x}^{\overline x}\frac{1}{x} dx=\ln\overline x-\ln\underline x
\label{6.10}
\ee

We now take the limit $N\rightarrow -\infty$, $M\rightarrow \infty$:

\ba
\int_{-\infty}^{2M}f(x)&=&\lambda\sum_{\mu=-\infty}^M\bra{+,2\mu}Lxf(x)\ket{2\mu,+}\label{6.11}\\
\int^{\infty}_{2M}f(x)&=&\lambda\sum_{\mu=M+1}^\infty \bra{+,2\mu}Lxf(x)\ket{2\mu,+}\no
\ea
and

\be
\int_{-\infty}^\infty f(x)=\lambda\sum_{\mu=-\infty}^\infty\bra{+,2\mu}Lxf(x)\ket{2\mu,+}
\label{6.12}
\ee
The factor $x$ in the matrix element allows the sum to
converge for $N\rightarrow -\infty$ for fields that do not vanish at $x=0$.

Similar formulas are obtained for $\mu, M$ odd.

The case $\sigma=-1$ can be treated analogously. From
the discussions on selfadjoint operators in the previous 
chapter we know that $\sigma=+1$ and $\sigma=-1$ should be
considered simultaneously.

\bigskip
\pn
\underline{The Hilbert space $L^2{}_q$}

The integral (\ref{6.12}) can be used to define a scalar
product for fields. We shall assume that the integral 
exists for even and odd values of $\mu$ and for
$\sigma=+1$ and $\sigma=-1$.

We define the scalar product  for $L^2{}_q$:

\be
(\chi,\psi)=\int \overline\chi\psi=\frac{\lambda}{2}
\sum_{\mu=-\infty \atop \sigma=+1,-1}^{\infty}\sigma\bra{\sigma,\mu}Lx\:\overline\chi\:\psi\ket{\mu, \sigma}
\label{6.13}
\ee

For fields that vanish at $x=\pm\infty$ we conclude from  (\ref{6.2}) that

\be
\int \nabla(\overline\chi \:\psi)=0
\label{6.14}
\ee
This leads to a formula for partial integration:

\ba
\int(\nabla\overline\chi)(L\:\psi)+\int(L^{-1}\overline\chi)(\nabla\psi)&=&0\label{6.15}\\
\int(\nabla\overline\chi)(L^{-1}\psi)+\int(L\:\overline\chi)(\nabla\psi)&=&0
\no
\ea

From Green's theorem follows:

\be
\int (\nabla^2\overline\chi)\:\psi=\int \overline\chi\:(\nabla^2\psi)
\label{6.16}
\ee

This shows that $\nabla^2$ is a hermitean operator in
$L^2{}_q$.

Eigenfunctions of the operator have been obtained in (\ref{4.20})
These are the functions $cos_q(kx)$ and $sin_q(kx)$. From the discussion of these
functions in chapter four we know that $kx$ has to be an even power of
$q$ for the functions to be in $L^2{}_q$. We therefore split
$L^2{}_q$ in four subspaces with $\mu$ even or $\mu$ odd and
$\sigma=+1$ or $\sigma=-1$.

\be
L^2{}_q={\cal H}_{\sigma=+1}^{\mu \:even}\oplus{\cal H}_{\sigma=+1}^{\mu \:odd}\oplus{\cal H}_{\sigma=-1}^{\mu \:even}\oplus{\cal H}_{\sigma=-1}^{\mu \:odd}
\label{6.17}
\ee

The functions

\be
N_q\left(\frac{2q}{\lambda}^\frac{1}{2}\right)^\frac{1}{2} q^k \cos_q(xq^{2k+1})
\label{6.18}
\ee
form an orthogonal basis in ${\cal H}_{\sigma=+1}^{\mu\:even}$. 
This follows from (\ref{4.11}) and the definition
of the scalar product (\ref{6.13}) has to be remembered. The
function (\ref{6.18}) belongs to the eigenvalue:

\be
\nabla^2 \cos_q(xq^{2k+1})=-\frac{1}{\lambda^2}q^{4k+1}\cos_q(xq^{2k+1})
\ee

We now give a table for the various eigenfunctions:

\vspace{4mm}
\begin{center}
\begin{tabular}{cc|c}
&Function&Eigenvalue\\${\cal H}_{\sigma=+1}^{even}:$&&\\ &$N_q\sqrt{\frac{2q}{\lambda}}q^k\cos_q(xq^{2k+1})$&$-\frac{1}{\lambda^2}q^{4k+1}$\\&&\\ \hline &&\\
&$N_q\sqrt{\frac{2q}{\lambda}}q^k\sin_q(xq^{2k+1})$&$-\frac{1}{\lambda^2}q^{4k+3}$\\&&\\ \hline &&\\
${\cal H}_{\sigma=+1}^{odd}:$&\\&$N_q\sqrt{\frac{2q}{\lambda}}q^k\cos_q(xq^{2k})$&
$-\frac{1}{\lambda^2}q^{4k-1}$\\&&\\\hline&&\\
&$N_q\sqrt{\frac{2q}{\lambda}}q^k\sin_q(xq^{2k})$&$-\frac{1}{\lambda^2}q^{4k+1}$
\end{tabular}
\end{center}
\vspace{4mm}

For $\sigma=-1$ we obtain the same table. The $cos_q$ functions
as well as the $sin_q$ functions form an orthonormal basis in the
respective subspaces of $L^2{}_q$. The set of eigenfunctions
is overcomplete. We conclude that $\nabla^2$ is hermitean but
not selfadjoint. On any of the bases defined above a selfadjoint
extension of the operator $\nabla^2$ can be defined. The
set of eigenvalues depends on the representation.


\section{Variational principle}\label{sub7}

\bigskip
\pn
\underline{Euler-Lagrange equations:}

We assume that there is a Lagrangian that depends on the fields
$\psi$, $\nabla L^{-1}\psi$ and $\dot{\psi}$. An action is defined

\begin{equation}
W=\int_{t_1}^{t_2} dt\int {\cal L} (\dot{\psi},\psi , \nabla L^{-1} \psi )
\label{7.1}
\end{equation}
In this chapter the integrals over the algebra are taken from 
$-\infty$ to $+\infty$ if not specified differently.

The action should be stable under the variation of the fields as they
are kept fixed at $t_1$ and $t_2$

\begin{equation}
\label{7.2}
\delta \psi (t_1)=\delta\psi (t_2)=0
\end{equation}

The variation of the action is:

\begin{eqnarray}
\delta W&=&\int_{t_1}^{t_2} dt \int \left\{ \frac{\partial {\cal L}}{\partial \dot{\psi}} \delta \dot{\psi}+ \frac{\partial {\cal L}}{\partial \psi}\delta \psi +\frac{\partial {\cal L}}{\partial\nabla L^{-1}\psi} \delta \nabla L^{-1}\psi \right\} \label{7.3}\\
&=&\int_{t_1}^{t_2} dt \int \left\{ -\frac{d}{dt} \frac{\partial {\cal L}}{\partial \dot{\psi}} - \nabla L \frac{\partial {\cal L}}{\partial\nabla L^{-1}\psi} + \frac{\partial {\cal L}}{\partial \psi} \right\} \delta\psi
\no
\end{eqnarray}

The boundary terms do not contribute because of  (\ref{7.2})
and $\psi$ and its variation  are supposed to vanish at infinity.
The partial integration for $\nabla L$ follows from
(\ref{1.27}) when we use it for $(Lf)(L^{-1}g)$:

\begin{equation}
\nabla \left( (Lf)(L^{-1}g) \right) =(\nabla Lf)(g)+f(\nabla L^{-1}g)
\label{7.4}
\end{equation}

We obtain the Euler-Lagrange equations:

\begin{equation}
\frac{\partial {\cal L}}{\partial \psi}=\frac{d}{dt} \frac{\partial {\cal L}}{\partial \dot{\psi}}+ \nabla L \frac{\partial {\cal L}}{\partial\nabla L^{-1}\psi}
\label{7.5}
\end{equation}

The Schroedinger equation (\ref{2.1}) can be obtained from the following 
Lagrangian. The variations of $\psi$ and $\overline{\psi}$ are independent:

\begin{equation}
\label{7.6}
W=\int_{t_1}^{t_2}dt \int \left\{i\overline{\psi}\frac{\partial}{\partial t} \psi -\frac{1}{2mq}(\nabla L^{-1} \overline{\psi})(\nabla L^{-1} \psi) - \overline{\psi}V\psi \right\}
\end{equation}

\bigskip
\pn
\underline{Noether theorem:}

We study the variation of the action under an arbitrary variation
of the fields  and the time:
\begin{eqnarray}
t'&=&t+\tau\label{7.7}\\
\psi '(t')&=&\psi (t)+\Delta\psi
\no
\end{eqnarray}

The variation of the action is defined as usual:

\begin{equation}
\Delta W=\int_{t_1'}^{t_2'} dt'\int_{2N}^{2M} {\cal L}(\dot{\psi}'(t'),\psi '(t'),\nabla L^{-1} \psi '(t'))-\int_{t_1}^{t_2}dt\int {\cal L}
\label{7.8}
\end{equation}

First we change the $t'$ variable of integration to $t$.

\[
  dt'=dt(1+\dot{\tau})
\]
\begin{eqnarray}
\label{7.9}
  &&\int_{t_1'}^{t_2'}  dt'\int_{2N}^{2M} {\cal L}(\dot{\psi}'(t'),\psi '(t'),\nabla L^{-1} \psi '(t')) \\
  &=&\int_{t_1}^{t_2} dt(1+\dot{\tau}) \int_{2N}^{2M}  {\cal L}(\dot{\psi}'(t+\tau)),\psi '(t+\tau),\nabla L^{-1} \psi '(t+\tau))
\no
\end{eqnarray}

Next we make a Taylor expansion of ${\cal L}$ in time and obtain:

\begin{eqnarray}
  \delta W &=& \int_{t_1}^{t_2} \int_{2N}^{2M} \frac{d}{dt} \tau {\cal L} 
\label{7.10}\\
  &\quad& +\int_{t_1}^{t_2} dt \int_{2N}^{2M} {\cal L}(\dot{\psi}'(t),\psi '(t),\nabla L^{-1} \psi '(t))-{\cal L}
\no
\end{eqnarray}

The variation of $\psi$ at the same time we call $\delta \psi$:

\begin{equation}
\delta \psi = \Delta \psi - \tau \dot{\psi}
\label{7.11}
\end{equation}

and we obtain in the same way as we derived the Euler-Lagrange 
equations:

\begin{eqnarray}
\Delta W&=&\int_{t_1}^{t_2} dt \frac{d}{dt} \tau  \int_{2N}^{2M} {\cal L} \no\label{7.12}\\ 
&&+\int_{t_1}^{t_2} dt \int_{2N}^{2M} \left\{-\frac{d}{dt} \frac{\partial {\cal L}}{\partial \dot{\psi}}-\nabla L \frac{\partial {\cal L}}{\partial \nabla L^{-1}\psi} + \frac{\partial {\cal L}}{\partial \psi} \right\}\delta \psi \\
&&+\int_{t_1}^{t_2} dt  \int_{2N}^{2M}\left[ \frac{d}{dt} \left\{  \frac{\partial {\cal L}}{\partial \dot{\psi}}\delta \psi \right\} + \nabla \left\{ \left(L \frac{\partial {\cal L}}{\partial \nabla L^{-1}\psi}\right)(L^{-1}\delta \psi ) \right\}\right]
\no
\end{eqnarray}

As a consequence of the Euler-Lagrange equation 
we find
\begin{eqnarray}
\Delta W&=&\int_{t_1}^{t_2} dt \frac{d}{dt} \tau  \int_{2N}^{2M}\left( {\cal L}-\frac{\partial {\cal L}}{\partial \dot{\psi}} \dot{\psi}\right)\no \\ 
&&-\int_{t_1}^{t_2} dt \tau \int_{2N}^{2M} \nabla \left(\left(L \frac{\partial {\cal L}}{\partial \nabla L^{-1}\psi}\right)(L^{-1}\dot{\psi}) \right) 
\label{7.13}\\
&&+\int_{t_1}^{t_2} dt  \int_{2N}^{2M}\left\{ \frac{d}{dt} \left(  \frac{\partial {\cal L}}{\partial \dot{\psi}}\Delta \psi \right) + \nabla \left( \left(L \frac{\partial {\cal L}}{\partial \nabla L^{-1}\psi}\right)L^{-1}\Delta \psi \right) \right\}\no
\end{eqnarray}

If we know that $\Delta W=0$ because (\ref{7.9}) are symmetry transformations
of the action we obtain the Noether theorem

\begin{eqnarray}
\label{7.14}
 &&\frac{d}{dt} \left\{ \tau \left( {\cal L}-\frac{\partial {\cal L}}{\partial \dot{\psi}} \dot{\psi}\right)+ \frac{\partial {\cal L}}{\partial \dot{\psi}}\Delta \psi \right\} \\
&+& \nabla \left\{ \left(L \frac{\partial {\cal L}}{\partial \nabla L^{-1}\psi}\right)\left(L^{-1}\Delta \psi  -\tau L^{-1} \dot{\psi}\right)\right\}=0
\no
\end{eqnarray}

This equation holds for the densities because the limits of the
integrations are arbitrary. 

\bigskip
\pn
\underline{Charge conservation:}

The Lagrangian  (\ref{7.6}) is invariant under phase transformations:
\begin{equation}
\psi '=e^{i\alpha}\psi , \quad \tau =0 ,\quad \Delta\psi =\delta\psi=i\alpha\psi 
\label{7.15}
\end{equation}

For the charge density we find from (\ref{7.14}) and (\ref{7.6})

\begin{eqnarray}
-\alpha\rho = \frac{\partial {\cal L}}{\partial \dot{\psi}}\delta \psi +\frac{\partial {\cal L}}{\partial \dot{\overline{\psi}}}\delta \overline{\psi} &=&-\alpha\overline{\psi}\psi\label{7.16}\\
&=& -\alpha\rho
\no
\end{eqnarray}

For the current density

\begin{eqnarray}
\left(L\frac{\partial {\cal L}}{\partial \nabla L^{-1}\psi}\right)(L^{-1}\delta \psi)&+&\left(L\frac{\partial {\cal L}}{\partial \nabla L^{-1}\overline{\psi}}\right)(L^{-1}\Delta \overline{\psi})\no\\
&=&-\alpha \frac{1}{2im}L^{-1}\left\{\overline{\psi}(L\nabla \psi)-(L\nabla \overline{\psi})\psi \right\} \label{7.17}\\
&=&-\alpha j
\no
\end{eqnarray}

This agrees with the definition of $\rho$ and $j$ in eqn (\ref{2.13}).

\bigskip
\pn
\underline{Energy conservation:}

The action (\ref{7.6}) is invariant under time translations:

\begin{eqnarray}
t'&=&t+\tau _0 \label{7.18}\\
\Delta\psi &=&0
\no
\end{eqnarray}

From (\ref{7.14}) we find the energy density and the density of the 
energy flow:

\begin{eqnarray}
{\cal H}&=&\frac{1}{2mq}(\nabla L^{-1}\overline{\psi})(\nabla L^{-1}\psi)+V\overline{\psi}\psi \label{7.19}\\
\pi &=&\frac{1}{2m}\left\{(\nabla\overline{\psi})L^{-1}\dot{\psi}+(L^{-1}\dot{\overline{\psi}}\nabla\psi \right\}
\no
\end{eqnarray}

\section{The Hilbert space $\L^2{}_q$}\label{sub8}

It is easy to find a representation of the algebra (\ref{1.1})
in the Hilbert space $L^2$ of square integrable functions. The usual
quantum mechanical operators in this space will be denoted with a hat:

\begin{equation}
[\hat{x},\hat{p}]=i,\:\: \overline{\hat x}=\hat x, \:\:\overline{\hat p}=\hat p
\label{8.1}
\ee

A further operator that we will use frequently is:

\be
\hat z=-\frac{i}{2}(\hat x\hat p +\hat p\hat x), \:\:\overline{\hat z}=-\hat z
\label{8.2}
\ee
It has the commutation property:

\be
\hat x \hat z=(\hat z +1)\hat x, \:\: \hat p \hat z=(\hat z -1)\hat p
\label{8.3}
\ee

An immediate consequence of (\ref{8.3}) is that the algebra (\ref{5.1})
is represented by

\be
x=\hat x,\:\:\Lambda=q^{\hat z}
\label{8.4}
\ee
$x$ is selfadjoint and $\Lambda$ is unitary.

The element $p$ can be expressed by the formula (\ref{1.7})

\be
p=i\lambda^{-1}\left(q^{-\frac{1}{2}}q^{\hat z}-q^{\frac{1}{2}}q^{-\hat z}\right){\hat x}^{-1}
\label{8.5}
\ee

That this represents  the algebra (\ref{1.1}) can be easily verified 
with the help of (\ref{8.1}). It is a consequence of the algebraic 
properties of $\hat{x}$ and $\hat{p}$ only. Any unitary transformation
of $\hat{x}$ and $\hat{p}$ will not change this property.

A short calculation going through the steps:

\be
\hat z=-i\hat x\hat p -\frac{1}{2},\:\: i{\hat x}^{-1}=\left(\hat z-\frac{1}{2}\right)^{-1}\,\hat p
\label{8.6}
\ee
leads to the following representation for $p$:

\be
p=\frac{[\hat z-\frac{1}{2}]}{\hat z-\frac{1}{2}}\hat p
\label{8.7}
\ee
where we use the symbol $[A]$ 

\be
[A]=\lambda^{-1}(q^A-q^{-A})
\label{8.8}
\ee
for any $A$.

It can be shown that $p$ as defined in (\ref{8.7}) is hermitean but not 
a selfadjoint operator in $L^2$.

We now show that it is possible to define a Hilbert space where $p$
as defined in (\ref{8.5}) is selfadjoint. We start from functions 
$f(\hat x)$ that are of class $C^\infty$ on $|\hat{x}|>0$.
They form the algebra of $C^\infty$ functions which we denote by $F$.

The set of functions $f(\hat{x})$ that vanish at 
$\hat x=\pm x_0 q^n, \:n \in \zet$
forms an ideal in $F$. We shall denote this ideal by 
$F_{\pm q^n}^{x_0}$ and introduce the factor space:

\be
{\cal F}_{x_0}^q=F \big/ (F_{+q^n}^{x_0}\cap F_{-q^n}^{x_0})
\label{8.9}
\ee

We will show that the operators $x,\Lambda$ and $p$ as defined by
(\ref{8.4}) and (\ref{8.5}) act on this factor space, that it is
possible to introduce a norm on ${\cal F}_{x_0}^q$,
defining a Hilbert space this way and that $x$ and $p$ are represented
by selfadjoint operators in this Hilbert space and that $\Lambda$ is
unitary.

We have to show that $x$ and $\Lambda$ leave the ideals
$F_{\pm q^n}$ invariant. For $x$ this is obvious, to show it for $\Lambda$
we use (\ref{5.1}) and find

\be
\Lambda f(\hat x)=q^{-\frac{1}{2}}f(\frac{1}{q} \hat x)
\label{8.10}
\ee

Here we consider $\Lambda$ to be represented by (\ref{8.4}) 
acting on $f\in F$. Thus $\Lambda$ transforms the zeros of an element of 
$F_{q^n}$ into the zeros of the transformed element and this leaves the
ideal invariant. The operator $p$ as represented by (\ref{8.5})
or (\ref{8.7}) is entirely expressed in terms of $x$ and $\Lambda$
and thus leaves the ideal invariant as well.

The algebra 
(\ref{1.1}) is now represented on ${\cal F}_{x_0}^q$.

Next we show that it is possible to define a scalar product on 
${\cal F}_{x_0}^q$. Guided by (\ref{6.13}) we define:

\be
(g,f)=\lambda\sum_{n=-\infty \atop{\sigma=+1,-1}}^{\infty} q^n \,\overline g(\sigma q^n x_0)f(\sigma q^n x_0)
\label{8.11}
\ee

This is the well known Jackson integral. It defines a norm on 
${\cal F}_{x_0}^q$ because the scalar product (\ref{8.11}) for a function of
$F_{\pm q^n}^{x_0}$ and any function of $F$ is zero.

The operators $x,\Lambda$ and $p$ are now linear operators in 
this Hilbert space. For $x$ it is obvious that it is selfadjoint
as it is diagonal. Let us show that $\Lambda$ is unitary:

\ba
\Lambda f(x_0 q^n)&=&q^{-\frac{1}{2}}f(x_0 q^{n-1}), \quad 
\:\Lambda^{-1}f(x_0 q^n)=q^{\frac{1}{2}}f(x_0 q^{n+1})\label{8.12}\\
(\Lambda g,\Lambda f)&=&\lambda\sum_{n=-\infty
\atop{\sigma=+1,-1}}^{\infty}
q^n q^{-1} \, \overline g (q^{n-1} x_0)f(q^{n-1} x_0) \nonumber \\
&=&(g,f)\label{8.13}
\ea

We now turn to the operator $p$. We use the expression (\ref{1.7})
for $p$:

\be
p = i \lambda^{-1} x^{-1} \left\{q^{\frac{1}{2}} \Lambda-q^{-\frac{1}{2}}
\Lambda^{-1}\right\}
\label{8.14}
\ee

and have it act on elements of ${\cal F}$. Using (\ref{8.12})
we find:

\be
p f(x)= i \lambda^{-1} x^{-1} \left\{f(\frac{1}{q}x)-f(qx)\right\}
\label{8.15}
\ee

The elements $f\in F$ are considered as representants of the cosets
in ${\cal F}_{x_0}^q$ such that (\ref{8.14}) is derived by acting with the differential 
operator $p$ on elements of $F$.

We know that $sin_q(x)$ and $cos_q(x)$ solve the eigenvalue equations of $p$ as defined
in (\ref{8.20}). this is just equation (\ref{4.3}). This equation links even
(odd) powers of $q$ in the argument of $cos_q(x)$ or $sin_q(x)$ to odd (even) powers. But
we know from chapter 4 that the behaviour of $cos_q(q^{2n+1})$ and
$sin_q(q^{2n+1})$ for $n \rightarrow \infty$ does not allow these functions to be in $L^2$ or in the
Hilbert space defined with the norm (\ref{8.11}).

To show that $p$ has a selfadjoint extension we shall draw from our
experience in chapter five. There it was essential to split states 
with eigenvalues belonging to even and odd powers of $q$. In 
chapter six  (\ref{6.17}) we have made the same experience.

We shall  start from four ideals in $F$: $F_{+q^{2n}},\:F_{+q^{2n+1}},\:F_{-q^{2n}}$ and $F_{-q^{2n+1}}$
For $x_0$ we again choose $x_0=1$ and drop it in the notation.
The respective factor spaces we denote by ${\cal F}_+^{2n},\:{\cal F}_+^{2n+1},\:{\cal F}_-^{2n},\:$ and ${\cal F}_-^{2n+1}$.

The intersection of the ideals leads to the union of the 
factor spaces and we identify ${\cal F}_1^q$ with this union.

The operator $x$ acts on each of these factor spaces individually. The
operator $\Lambda$ maps ${\cal F}_{+(-)}^{2n}$ onto ${\cal F}_{+(-)}^{2n+1}$ 
and vice versa. This follows from  (\ref{8.12}) which shows that the
ideals $F_{\pm q^{2n}}(F_{\pm q^{2n+1}})$
are mapped to the ideals $F_{\pm q^{2n+1}}(F_{\pm q^{2n}})$.

We shall distinguish functions $f\in F$ as representants 
of ${\cal F}_{\pm}^{2n}$ and ${\cal F}_{\pm}^{2n+1}$:

\be
f^g\in {\cal F}_{\pm}^{2n},\: f^u\in{\cal F}_{\pm}^{2n+1}
\label{8.16}
\ee

Thus we have to read (\ref{8.11})
as follows:
\be
\Lambda f^{g(u)}(x)=\Lambda f(x)=q^{-\frac{1}{2}}f(\frac{1}{q}x)=q^{-\frac{1}{2}}f^{u(g)}(\frac{1}{q}x)\label{8.17}
\ee

The same way we have to read (\ref{8.15}):

\be
p\, f^{g(u)}(x)= i \lambda^{-1} x^{-1} \left\{f^{u(g)}(\frac{1}{q}x)-f^{u(g)}(qx)\right\}\label{8.18}
\ee

To show that $p$ has a selfadjoint extension we construct a 
basis where $p$ is diagonal. Eqns. (\ref{8.17}) and (\ref{4.3})
show that $cos_q(x)$ and $sin_q(x)$ will play the role of transition functions.

We first choose specific  representatives in the spaces ${\cal F}_+^{2n}$ and
${\cal F}_+^{2n+1}$ 

\ba
{\cal C}_+^g(q^{2n})^{(\nu)}&\sim&\cos_q(xq^{2\nu})\no\\
{\cal C}_+^u(q^{2n+1})^{(\nu)}&\sim&\cos_q(xq^{2\nu+1})\label{8.19}\\
{\cal S}_+^g(q^{2n})^{(\nu)}&\sim&\sin_q(xq^{2\nu})\no\\
{\cal S}_+^u(q^{2n+1})^{(\nu)}&\sim&\sin_q(xq^{2\nu+1})\no
\ea

We calculate according to the rule (\ref{8.18}):

\ba
p\:{\cal C}_+^{g(\nu)}&\sim& p\cos_q(xq^{2\nu})=i\lambda^{-1} x^{-1}\left\{\cos_q(xq^{2\nu-1})-\cos_q(xq^{2\nu+1})\right\}\no\\
&=&i\lambda^{-1}\frac{1}{q}\sin_q(q^{-1}x)\no\\
&=&i\lambda^{-1}q^{2\nu-1}\sin_q(xq^{2\nu-1})\no\\
&=&i\lambda^{-1}q^{2\nu-1}{\cal S}_+^{u(\nu-1)}\label{8.20}
\ea

A similar calculation shows that:

\be
p\:{\cal S}_+^{u(\nu-1)}=-i\lambda^{-1}q^{2\nu}{\cal C}_+^{g(\nu)}\label{8.21}
\ee

The map $p$ can now be diagonalized in the space ${\cal F}_+^{2n}\cup{\cal F}_+^{2n+1}$.
A set of eigenvectors are:
\be
{\cal C}_+^{g(\nu)}\mp iq^{-\frac{1}{2}}{\cal S}_+^{u(\nu-1)}\in{\cal F}_+^{2n}\cup{\cal F}_+^{2n+1}\label{8.22}
\ee
They belong to the eigenvalues:

\be
p\left \{{\cal C}_+^{g(\nu)}\mp iq^{-\frac{1}{2}}{\cal S}_+^{u(\nu-1)}\right\}=\pm \frac{1}{\lambda q^{\frac{1}{2}}} q^{2\nu}\left\{{\cal C}_+^{g(\nu)}\mp iq^{-\frac{1}{2}}{\cal S}_+^{u(\nu-1)}\right\}\label{8.23}
\ee

When we compare (\ref{8.23}) with (\ref{5.18}) we know how
to continue. It is exactly the same analysis as in chapter five that
we have to repeat to construct a basis that consists of eigenstates of
$p$.

We now turn our attention to the operator $p^2$. As a differential
operator it acts on $C^\infty$ functions. Starting from the representation
(\ref{8.7}) for $p$  a short calculation yields:

\ba
p^2&=&\frac{4}{\lambda^2}\,\hat p\: \frac{q+q^{-1}-2\cos(\hat x \hat p +\hat p \hat x)h}{1+(\hat x \hat p +\hat p \hat x)}\,\hat p\no\\
&=&{\hat p}^2+O(h)\label{8.24}\\
q&=&e^h.\no
\ea

We can consider $h$ as a coupling constant and expand $p^2$ at 
$\hat{p}^2$. Thus we could consider a Hamiltonian:

\be
H=\frac{1}{2}\,p^2=\frac{1}{2}\,{\hat p}^2+O(h)\label{8.25}
\ee

We would find that $H$ is not a selfadjoint operator in $L^2$. 
It is, however, a selfadjoint operator on the space ${\cal F}^q$ equipped 
with the norm (\ref{8.11}). Its eigenvalues are $\frac{1}{2q\lambda}q^{4\nu}$ and
$\frac{1}{2q\lambda}q^{4\nu+2}$. The eigenstates are the same as the eigenstates of $p$.

A final remark on the representation of $p$ and $\Lambda$ on the space
${\cal F}^q$. Both can be viewed as a linear mapping of ${\cal F}^q$ into ${\cal F}^q$
in very much the same way as a vector field is a linear mapping
of $C^{\infty}$ into $C^{\infty}$. The derivative property of a vector field is changed, 
however. The operator $\Lambda$ acts group like:
\be
\Lambda fg=(\Lambda f)(\Lambda g)
\ee
and $p$ acts with the Leibniz rule:

\be
pfg=(pf)(\Lambda g)+(\Lambda ^{-1})(pg)
\ee

They form an algebra:

\be
(\Lambda p-qp\Lambda)f=0
\ee

\section{Gauge theories on the factor spaces}\label{sub9}

The usual gauge transformations on wave functions $\psi(\hat x,t)$

\ba
\psi '(\hat x,t)&=&e^{i\alpha(\hat x,t)}\psi(\hat x,t)\label{9.1}\\
\alpha(\hat x,t)&=&g\sum_l\alpha_l(\hat x,t)T_l\no 
\ea
define a gauge transformation on ${\cal F}^q$ as well. We
want to show that there are covariant expressions for
$x, \Lambda$ and $p$ that reduce to $x, \Lambda$ and $p$
for $g=0$.

The operator $x$ by itself is obviously covariant and acts on ${\cal F}^q$.

A covariant expression of $\Lambda$ is obtained by
replacing the canonical momentum in (\ref{8.4}) by the covariant
momentum

\be
\hat p \rightarrow \hat p-gA^l(\hat x,t)T_l\label{9.2}
\ee
where $A^lT_l$ transforms in the usual way:
\be
A'^lT_l=e^{i\alpha}A^lT_l e^{-i\alpha}-\frac{i}{g}e^{i\alpha}\, \partial\, e^{-i\alpha}\label{9.3}
\ee
This leads to the expression

\be
\tilde \Lambda=q^{-i\hat x(\hat p-gA^lT_l)-\frac{1}{2}}\label{9.4}
\ee
It is certainly true that

\be
\widetilde \Lambda'=e^{i\alpha(\hat x,t)}\Lambda \,e^{-i\alpha(\hat x,t)}\label{9.5}
\ee
because it is true for any power of the covariant derivative
$\hat p -g A^lT_l$. Thus:

\be
\widetilde \Lambda' \psi'(\hat x,t)=e^{i\alpha(\hat x,t)}\widetilde \Lambda \,\psi(\hat x,t)\label{9.6}
\ee

We have to show that $\widetilde\Lambda$ is defined on ${\cal F}^q$. This will be the case if
$\widetilde \Lambda$ leaves the ideals $F_{\pm q^n}$ invariant. We adapt the
notation of (\ref{3.19}) and define:

\ba
E^{-1}&=&q^{i\hat x \hat p +\frac{1}{2}}\widetilde\Lambda=q^{i\hat x \hat p}q^{-i\hat x(\hat p - gA^lT_l)}\label{9.7}\\
E&=&q^{i\hat x(\hat p - gA^lT_l)}q^{-i\hat x \hat p}
\no
\ea

As a consequence of the Baker-Campbell-Hausdorff formula $E$ can
be written in the form

\be
E=q^{ia_lT^l}\label{9.8}
\ee
where $a$ is a functional of $A$ and its space derivatives. Thus
$E$ acts on ${\cal F}_q$.

$E$ was defined such that 

\be
\widetilde\Lambda=\Lambda E^{-1}(\hat x)\label{9.9}
\ee
This shows that $\tilde\Lambda$ acts on ${\cal F}_q$ as $E^{-1}$ and $\Lambda$ do so.

Formula (\ref{9.7}) can be used to compute $a$ in (\ref{9.8})
explicitely. We make use of the time dependence in $A(\hat x,t)$
and compute the expression $E\frac{\partial}{\partial t}E^{-1}$
first with $E$ in the form (\ref{9.8})
and then with $E$ in the form (\ref{9.7}) and compare the result.
We demonstrate this for the abelian case:

\be
E\frac{\partial}{\partial t}E^{-1}=\frac{\partial}{\partial t}-i\frac{\partial}{\partial t}ah,\quad q=e^h\label{9.10}
\ee
Now the second way:

\ba
E\frac{\partial}{\partial t}E^{-1}&=&q^{i\hat x(\hat p-gA)}\frac{\partial}{\partial t}q^{-i\hat x(\hat p-gA)}\label{9.11}\\
&=&\frac{\partial}{\partial t}+\left[-ig\hat xAh,\frac{\partial}{\partial t}\right]+\frac{1}{2}\left[i\hat x(\hat p-gA)h,\left[-ig\hat x Ah,\frac{\partial}{\partial t}\right]\right]+\ldots\no\\
&=&\frac{\partial}{\partial t}+igh\hat x\dot A+\frac{1}{2}h^2g\hat x\frac{\partial}{\partial \hat x}(ig\hat x\dot A)+ \ldots\no\\
&&+\frac{1}{n!}h^n(\hat x\frac{\partial}{\partial \hat x})^{n-1}(ig\hat x\dot A) + \ldots\no
\ea
This can be written in a more compact way:

\be
E\frac{\partial}{\partial t}E^{-1}=\frac{\partial}{\partial t}+\frac{q^{\hat x\frac{\partial}{\partial \hat x}}-1}{\hat x\frac{\partial}{\partial \hat x}}ig\frac{\partial}{\partial t}\hat xA\label{9.12}
\ee

A comparison of (\ref{9.12}) and (\ref{9.10}) shows that

\be
a=-g\frac{q^{\hat x\frac{\partial}{\partial \hat x}}-1}{h\hat x\frac{\partial}{\partial \hat x}}\hat x A\label{9.13}
\ee
In an expansion in $\hat x\frac{\partial}{\partial \hat x}$ of 
(\ref{9.13}) there is never a negative power. Another way of writing
(\ref{9.13}) is:

\ba
a&=&-g\frac{q^{-\frac{1}{2}}\Lambda^{-1}-1}{h\hat x\frac{\partial}{\partial \hat x}}\hat x A\label{9.14}\\
&=&-g h^{-1}(q^{-\frac{1}{2}}\Lambda^{-1}-1)\partial_x^{-1}A(\hat x,t)
\no
\ea
In this form it is obvious that $a$ has the right transformation
property such that $E$ transforms as in (\ref{3.17}).

We show the transformation law in the non-abelian case for $E$
as well:

\ba
E(\hat x)&=&q^{i\hat x(\hat p-gA^l(\hat x)T_l)}q^{-i\hat x\hat p}\label{9.15}\\
E'(\hat x)&=&q^{i\hat x(\hat p-gA'^l(\hat x)T_l)}q^{-i\hat x\hat p}\label{9.16}\\
&=&e^{i\alpha(\hat x)}q^{i\hat x(\hat p-gA^l(\hat x)T_l)}e^{-i\alpha(\hat x)}q^{-i\hat x\hat p}\no\\
&=&e^{i\alpha(\hat x)}E(\hat x)e^{-i\alpha(q\hat x)}\no
\ea
This agrees with (\ref{3.7})-

It is now obvious how to define the covariant version of $p$

\be
\tilde p=i\lambda^{-1}\hat x^{-1}(q^{\frac{1}{2}}\tilde \Lambda-q^{-\frac{1}{2}}\tilde \Lambda^{-1})\label{9.17}
\ee

This is covariant under gauge transformations, it has the property that
$\tilde p \rightarrow p$ for $g\rightarrow 0$ and it is defined on ${\cal F}_q$.

The gauge covariant differential operators $\tilde\Lambda,\: \tilde p$ and ${\cal D}_t$
acting on time dependent elements of ${\cal F}_q$ form an algebra. This 
is the algebra (\ref{3.50}). We can now express the tensor $T$
through the vector potential $A^1=A,\:-igA^0=\omega$

\be
T=-E(\partial _t E^{-1})-igA^0+igE(\Lambda^{-1}A^0)E^{-1}\label{9.18}
\ee
The first expression we have calculated in 
(\ref{9.11}) for an abelian gauge group. We now compute $T$ for an
abelian gauge group:

\be
T=\frac{1}{\hat x\frac{\partial}{\partial \hat x}}ig\left\{(\Lambda ^{-1}-1)\hat x\left(-\frac{\partial}{\partial t}A^1+\frac{\partial}{\partial \hat x}A^0\right)\right\}
\ee
We see that the curvature $T$ is a functional of the
usual curvature $F$:

\be
F=\frac{\partial}{\partial \hat x}A^0-\frac{\partial}{\partial t}A^1
\label{9.19}
\ee
and its space derivatives

\be
T=ig(\Lambda^{-1}-1)\partial_x^{-1}F
\label{9.21}
\ee

For the non-abelian case the calculations are more involved and as I 
have not done them I have to leave them to the reader.

\newpage
\chapter{$q$-Deformed Heisenberg algebra in $n$ dimensions}\label{Sekt2}
\section{$SL_q(2)$, Quantum groups and the $R$-matrix.}\label{sub2.1}
Let me first present a simple example of a 
quantum group, $SL_q(2)$. This will allow
me to exhibit the mathematical structure of quantum groups and at the
same time I can demonstrate why a physicist might get interested in
such an object.

Take a two by two matrix with entries $a,b,c,d$:
\be
T=\left(\begin{array}{cc}
 a&b\\ c&d\end{array} \right)
\label{qw1.1}\ee
If the entries are complex numbers or real numbers and if the
determinant is not zero $T$ will be an element of $GL(2,C)$ or
$GL(2,R)$.

For $GL_q(2)$ we demand that the entries $a,b,c,d$ are elements of
an algebra ${\cal A}$ which we define as follows: We take the free
associative algebra generated by $1,a,b,c,d$ and divide by the
ideal generated by the following relations:
\ba
ab &=&qba\nonumber\\
ac &=&qca\nonumber\\
ad &=&da +\lambda bc \label{qw1.2}\\
bc &=&cb\nonumber\\
bd &=&qdb\nonumber\\
cd &=&qdc\nonumber
\ea
$q$ is a complex number, $q\in {\mathbb{C}}, q\not= 0$ and $\lambda =q-q^{-1}$.\\
In the algebra $\A$ we allow formal power series.\\
A direct calculation shows that
\be
det_qT=ad-qbc \label{qw1.3}
\ee
is central, it commutes with $a,b,c$ and $d$. If $det_qT\not= 0$
we call $T$ an element of $GL_q(2)$, if $det_qT=1$, $T$ will be an
element of $SL_q(2)$.

So far for the definition of $SL_q(2)$. It looks quite
arbitrary; by what follows, however, it will become clear that the
relations (\ref{qw1.2}) have been chosen carefully such that the
algebra $\A$ has some very nontrivial properties.

The relations (\ref{qw1.2}) allow an ordering of the elements $a,b,c,d$. We could
decide to write any monomial of degree $n$ as a sum of monomials
$a^{n_1}b^{n_2}c^{n_3}d^{n_4}$ with $n=n_1+n_2+n_3+n_4$.
We could have chosen any other ordering, e.g.\
$b^{m_1}a^{m_2}d^{m_3}c^{m_4},  \quad n=m_1+m_2+m_3+m_4$. Moreover,
the monomials in a given order are a basis for polynomials of fixed
degree (Poincar\'e-Birkhoff-Witt).

The relations depend on a parameter $q$ and for $q=1$ the algebra
becomes commutative. In this sense we call the quantum group
$GL_q(2)$ a $q$ deformation of $GL(2,C)$.

That the Poincar\'e-Birkhoff-Witt property is far from
being trivial can be demonstrated by the following example. Consider
an algebra freely generated by two elements $x,y$ and divided
by the ideal generated by the relation $yx=xy+x^2+y^2$. If we now
try to order $y^2x$ in the $xy$ ordering we find $x^3+y^3+x^2y+xy^2=0$
The polynomials of third degree in the $x,y$ ordering are not independent.
This algebra does not have the Poincar\'e-Birkhoff-Witt 
property.\footnote{This example has been shown to me by Phung Ho Hai
(thesis).}

That our algebra $\A$ has the Poincar\'e-Birkhoff-Witt property follows
from the fact that it can be formulated with the help of an $\hat R$
matrix and that this matrix
satisfies the quantum Yang Baxter equation.

Let me introduce the concept of an $\hat R$ matrix. The relation
(\ref{qw1.2}) can be written in the form
\be
\hat R^{ij}{}_{kl} T^k{}_rT^l{}_s=T^i{}_kT^j{}_l \label{qw1.4}\\
\hat R^{kl}{}_{rs}
\ee
The indices take the value one and two, repeated indices are to be
summed, $T^i{}_j$ stands for $a,b,c,d$ in an obvious assignment
and $\hat R$ is the following 4 by 4 matrix
\be
\hat R=\left(\begin{array}{cccc}
                q&0&0&0\\
                0&\lambda &1&0\\
                0&1&0&0\\
                0&0&0&q\end{array}\right)\label{qw1.5}
\ee
The rows and columns of $\hat R$ are labelled by 11, 12, 21 and 22.

As an example:
\begin{displaymath}
\hat R^{12}{}_{ij}T^i{}_2 T^j{}_2=T^1{}_i T^2{}_j
\hat R^{ij}{}_{22}
\end{displaymath}
becomes
\begin{displaymath}
\lambda T^1{}_2T^2{}_2+T^2{}_2T^1{}_2 = qT^1{}_2T^2{}_2
\end{displaymath}
or
\begin{displaymath}
\lambda bd +db =qbd \rightarrow  bd=q db
\end{displaymath}
The relations (\ref{qw1.4}) are called $\hat R TT$ relations.
They are 16 relations that reduce to the 6 relations of eq.\ (\ref{qw1.2}).
This of course is due to particular properties of the $\hat R$ matrix
(\ref{qw1.5}). We could start from an arbitrary $\hat R$ matrix, but then
the $\hat R TT$ relations might have $T=1$ and $0$ as the only solutions. If
on the other hand we take $\hat R$ to be the unit matrix $\hat R^{ij}_{kl}=
\delta^i_k\delta^j_l$, there would be no relation for $abcd$.
If $\hat{R}^{ij}_{kl} = \delta^i_l \delta^j_k$ all elements of the $T$ matrix would commute.
The art is to find an $\hat R$ matrix that
guarantees the Poincar\'e Birkhoff Witt property, in this case for
polynomials in $a,b,c,d$.

In any case, the existence of an $\hat R$ matrix has far-reaching
consequences. E.g.\ from the $\hat R TT$ relations follows:
\begin{eqnarray*}
\lefteqn{\hat R^{i_1i_2}{}_{j_1j_2}(T^{j_1}{}_r\otimes T^r{}_{k_1})\cdot
(T^{j_2}{}_s\otimes T^s{}_{k_2})} \\
&=&\hat R^{i_1i_2}{}_{j_1j_2}T^{j_1}{}_rT^{j_2}{}_s\otimes T^r{}_{k_1}
  T^s{}_{k_2}\nonumber\\
&=&T^{i_1}{}_{j_1}T^{i_2}{}_{j_2}\hat R^{j_1j_2}{}_{rs}
\otimes T^r{}_{k_1}
  T^s{}_{k_2}\nonumber\\
&=&(T^{i_1}{}_r\otimes T^r{}_{l_1})\cdot(T^{i_2}{}_s\otimes
  T^s{}_{l_2})\hat R^{l_1l_2}{}_{k_1k_2}\nonumber
\end{eqnarray*}
This shows that we can define a co-product:
\be
\Delta T^j{}_l=T^j{}_r\otimes T^r{}_l
\label{qw1.6}
\ee
It is compatible with the $\hat R TT$
relations:

\be
\hat R^{ij}{}_{kl}\Delta T^k{}_r\Delta T^l{}_s=\Delta T^i{}_k
\Delta T^j{}_l\hat R^{kl}{}_{rs}\label{qw1.7}
\ee
This co-multiplication is an essential ingredient of a Hopf
algebra. Another one is the existence of an inverse (antipode). We have
already seen that $det_qT$ is central. We can enlarge the algebra by the
inverse of $det_qT$. Then it is easy to see that
\be
T^{-1}={1\over det_q T}\left(\begin{array}{cc}
                            d, &-{1\over q} b\\
                                -qc, &a\end{array}\right)\label{qw1.8}
\ee
is the inverse matrix of $T$.

We have now learned that a quantum group is a Hopf algebra, it is a
$q$-deformation of a group and it has an $\hat R$ matrix associated
with it.

We continue by studying the $\hat R$ matrix (\ref{qw1.5}) in more detail.
It is an easy exercise to verify that it satisfies the characteristic
equation:
\be
(\hat R-q)(\hat R+\frac{1}{q})=0
\label{qw1.9}
\ee
The matrix $\hat R$ has $q$ and $-\frac{1}{q}$ as eigenvalues.
The projectors that project on the respective eigenspaces follow
from the characteristic equation:
\be
\rm A = -\frac{q}{1+q^2}(\hat R-q)\qquad
S =\frac{q}{1+q^2}(\hat R+\frac{1}{q})
\label{qw1.10}
\ee
A is a deformation of an antisymmetrizxer and $S$ of a symmetrizer. The
normalization is such that:
\be
A^2=A,\quad S^2=S,\quad AS=SA=0.
\quad 1=S+A,\quad \hat R= qS-\frac{1}{q} A
\label{qw1.11}
\ee
The $\hat R$ matrix approach can be generalized to $n$ dimensions.
The $n^2$ by $n^2~\hat R$ matrix for the quantum group~$GL_q(n)$ is:
\be
\hat R^{ji}_{kl}=\delta^i_k\delta^j_l[1+(q-1)\delta^{ij}]+(q-\frac{1}{q})
\theta (i-j)\delta^j_k\delta^i_l \label{qw1.12}
\ee
where $\theta (i-j)=1\quad$ for $i>j$ and $\theta (i-j)=0\quad$ for
$i\leq j$.
The $\hat RTT$ relations (\ref{qw1.4})
now refer to an $n\times n$ matrix $T$. There are $n^4$ relations
for the $n^2$ entries of $T$. It can be shown that the thus defined
$T$ matrix has the Poincar\'e-Birkhoff-Witt property. Comultiplication
is defined the same way as by (\ref{qw1.6}) and the $\hat R$ matrix satisfies the
same characteristic equation (\ref{qw1.10}). This leads to the projectors
$A$ and $S$ in the $n$-dimensional case as well.

The $\hat R$-matrices (\ref{qw1.5}, \ref{qw1.12}) are symmetric:
\be
\hat R^{ab}{}_{cd}=\hat R^{cd}{}_{ab}
\label{qw1.13}
\ee
In such a case, the transposed matrix $\tilde T$:
\be
\tilde T^a{}_b=T^b{}_a
\label{qw1.14}
\ee
will also satisfy the RTT relations (\ref{qw1.4})
\ba
& &\tilde T^a{}_c\tilde T^b{}_d\hat{R}^{cd}{}_{ef}=
T^c{}_aT^d{}_b\hat{R}^{ef}{}_{cd}\nonumber\\
&=&T^e{}_iT^{f}{}_j\hat{R}^{ij}{}_{ab}=\hat R^{ab}{}_{ij}\tilde T^i{}_e
\tilde T^j{}_f
\label{qw1.15}
\ea
Therefore  $\tilde T\in GL_q (n)$, if $T\in GL_q (n)$.
It then has an inverse $\tilde T^{-1}$:
\be
\tilde T^{-1~a}{}_i\tilde T^i{}_b=
\delta^a_b,\/\qquad\tilde T^a{}_i\tilde T^{-1~i}{}_b
=\delta^a_b
\label{qw1.16}
\ee
In general, (for $q\not= 1$), however,
$\tilde T^{-1}\not=\widetilde{T^{-1}}.$
For $n=2$:
\be
\tilde T^{-1}=\frac{1}{det_q T}\left(\begin{array}{cc}
           d&-\frac{1}{q}c\\
          -qb &a\end{array}\right)\quad ,
\widetilde{T^{-1}}=\frac{1}{det_q T}\left(\begin{array}{cc}
           d&-q c\\
          -\frac{1}{q} b &a\end{array}\right).
\label{qw1.17}
\ee
\be
\left(\begin{array}{cc}
           q^2& 0\\
          0 &1\end{array}\right)\quad
\left(\begin{array}{cc}
           d&-\frac{1}{q} c \\
          -qb &a\end{array}\right)\quad
\left(\begin{array}{cc}
           {1 \over q^2} &0 \\
          0 &1\end{array}\right)\quad
= \left(\begin{array}{cc}
           d &-qc \\
       -\frac{1}{q} b &a\end{array}\right)\quad
\ee

\noindent
\section{Quantum planes}\label{sub2.2}

After having defined the algebraic structure of a quantum group we are 
interested in its comodules. Such comodules will be called  quantum
planes.

Let the matrix  $T$ satisfy $\hat{R}TT$ relations of the type (\ref{qw1.4}) with some 
general matrix $\hat{R}$ and let $T$ co-act 
on a $n$-component ``vector'' as follows:
\begin{equation}
\omega(x^i) = T^i{}_k \otimes x^k\label{qw2.1}
\end{equation}
This defines a contravariant vector.

We ask for an algebraic structure of the quantum plane that is
compatible with the co-action  (\ref{qw2.1}).

{From} the $\hat{R}TT$ relations follows that
for any polynomical ${\cal P}(\hat{R})$ it is true that:
\begin{equation}
{\cal P}(\hat{R})^{ij}{}_{kl} T^k{}_r T^l{}_s = T^i{}_k T^j{}_l {\cal P}(\hat{R})^{kl}{}_{rs}
\end{equation}
As a consequence, any relation of the type
\begin{equation}
{\cal P}(\hat{R})^{ij}{}_{kl}x^kx^l = 0\label{qw2.3}
\end{equation}
implies
\begin{equation}
{\cal P}(\hat{R})^{ij}{}_{kl}\omega(x^k)\omega(x^l)     = 0
\end{equation}
We shall say that the  relation (\ref{qw2.3}) is covariant. 

A natural definition of an algebraic structure is:
\begin{equation}
P_A^{ij}{}_{kl}x^kx^l = 0\label{qw2.5}
\end{equation}

if $P_A$, the ``antisymmetrizer'', can be obtained as a polynomial in the $\hat{R}$ matrix. This then generalizes 
the property that the coordinates of an ordinary space commute and (\ref{2.5})
is covariant.

For $GL_q(n)$ the most general polynomial of $\hat{R}$ is of degree one and from (\ref{1.10}) we know  that (\ref{qw2.5})
becomes:
\begin{equation}
x^ix^j = \frac{1}{q} \hat{R}^{ij}{}_{kl} x^kx^l\label{qw2.6}
\end{equation}
In two dimensions, this reduces to the condition:
\begin{equation}\label{qw2.7}
x^1x^2 = qx^2x^1
\end{equation}

The relations (\ref{qw2.6}) can be generalized to the case of two or more copies of quantum planes, for instance  $(x^1,x^2)$ and $(y^1,y^2)$.
The relations
\begin{equation}
x^iy^j = \frac{\kappa}{q} \hat{R}^{ij}{}_{kl} y^kx^l\label{qw2.8}
\end{equation}
are consistent, i.e.  they have the Poincar\'{e} -Birkhoff-Witt property for 
arbitrary $\kappa , \kappa \not=0$ and they are covariant. For $n=2$ the relations (\ref{qw2.8}) are
\[
x^1y^1 = \kappa y^1x^1
\]
\be
x^1y^2 = \frac{\kappa}{q} y^2x^1 + \frac{\kappa}{q} \lambda y^1x^2\label{qw2.9}
\ee
\[
x^2y^1 = \frac{\kappa}{q} y^1x^2
\]
\[
x^2y^2 = \kappa y^2x^2
\]
Consistency can be checked explicitely:
\begin{equation}
x^1(y^1y^2 - qy^2y^1) = \kappa y^1x^1y^2 - q\kappa(\frac{1}{q} y^2x^1 + \frac{1}{q} \lambda y^1x^2)y^1
= \kappa^2(\frac{1}{q} y^1y^2 - y^2y^1)x^1\label{qw2.10}
\end{equation}

Taking $x^1$ through the $y$ relations does not give rise to new relations. Similar for all third order 
relations and then, by induction, we conclude that (\ref{qw2.9}) does not generate new higher order relations and 
therefore the Poincar\'{e} -Birkhoff-Witt property holds.\\

We can do this more systematically if we start from a general $\hat{R}$
matrix as we did at the beginning of this chapter. We consider three copies of quantum planes, $x,y$ and $z$. Covariant relations are:
\begin{equation}\label{qw2.11}
xy = \hat{R}yx ,\quad  yz = \hat{R}zy ,\quad  xz = \hat{R}zx%
\end{equation}
(indices as in (\ref{qw2.8})). We demand that a reordering of $xyz$ to $zyx$ should give the same result, independent of the 
way we achieve this reordering. There are two ``independent'' ways to do it. The first one 
starts by changing first $xy$: $xyz \rightarrow yxz \rightarrow yzx \rightarrow zyx$, the 
second one by changing first $yz$: $xyz \rightarrow xzy \rightarrow zxy \rightarrow zyx$.

The result should be the same. This leads to an equation on the $\hat{R}$ matrix that is called 
Quantum Yang-Baxter
equation. It can easily be formulated by introducing $n^3$ by $n^3$ matrices:
\begin{equation}
\hat{R}_{12}{}^{(i)}{}_{(j)} = \hat{R}^{i_1 i_2}{}_{j_1j_2} \delta^{i_3}{}_{j_3}\label{qw2.12}
\end{equation}
and

\begin{equation}
\hat{R}_{23}{}^{(i)}{}_{(j)} = \delta^{i_1}{}_{j_1}\hat{R}^{i_2 i_3}{}_{j_2j_3}\label{qw2.13} 
\end{equation}
The Yang-Baxter equation that follows from the independence of the reordering then is:
\begin{equation}
\hat{R}_{12}\hat{R}_{23}\hat{R}_{12} = \hat{R}_{23}\hat{R}_{12}\hat{R}_{23}\label{qw2.14}
\end{equation}
These are $n^6$ equations for $n^4$ independent entries of the
 $\hat{R}$-matrix. It can be checked that the $\hat{R}$-matrices
 (\ref{qw1.5}) and (\ref{qw1.12}) do satisfy the Yang-Baxter equation.

A direct consequence of (\ref{qw2.14}) is
\be
{\cal P}(\hat{R}_{12}) \hat{R}_{23} \hat{R}_{12} =  \hat{R}_{23} \hat{R}_{12} {\cal P}(\hat{R}_{23})
\ee
where ${\cal P}(\hat{R}_{12})$ is any polynomial in $\hat{R}_{12}$, e.g. a projector $P_{12}$ or $\hat{R}_{12}{}^{-1}$.

We now use (\ref{qw2.14}) to discuss relations of the type  (\ref{qw2.10}) more systematically:
\[
P_{A23} xyy = \frac{\kappa}{q} P_{A23} \hat{R}_{12} yxy 
\]
\begin{eqnarray*}
&=& \frac{\kappa^2}{q^2} P_{A23} \hat{R}_{12} \hat{R}_{23} yyx =\\
&=& \frac{\kappa^2}{q^2} \hat{R}_{12} \hat{R}_{23} P_{A12} yyx   .
\end{eqnarray*}
Thus the relation (\ref{qw2.8}) is consistent with the $xx$ and $yy$ relations (\ref{qw2.3})
if $\hat{R}$ satisfies the Yang-Baxter equation.

For the $\hat{R}$-matrix (\ref{qw2.14}) becomes
\be
\hat{R}^{ab}{}_{\alpha \beta} \hat{R}^{\beta c}{} _{\gamma t}  \hat{R}^{\alpha \gamma}{}_{rs} =\label{qw2.16}
\hat{R}^{bc}{}_{\alpha \beta}\hat{R}^{a \alpha}{}_{r \gamma}  \hat{R}^{\gamma \beta}{}_{st}
\ee
It is interesting to note that this relation expresses the fact that the $n \times n$ matrices $t^\alpha{}_\gamma$
\be
(t^\alpha{}_\gamma)_{ab} = \hat{R}^{a \alpha}{}_{\gamma b}\label{qw2.17}
\ee
- where $a,b$ label the $n$ rows and $n$ columns respectively and $\alpha, \gamma$ label  $n^2$
matrices -  represent a solution of the $\hat{R}TT$ relations. With the notation (\ref{qw2.17}) eqn. (\ref{qw2.16}) takes 
the form
\be
t^b{}_\alpha t^c{}_\gamma \hat{R}^{\alpha \gamma}{}_{rs} =
R^{bc}{}_{\alpha \beta} t^\alpha{}_r t^\beta{}_s
\ee
where the $ts$ are multiplied matrixwise.

The inverse of $\hat{R}_{12}$ or $\hat{R}_{23}$ is $\hat{R}^{-1}{}_{12}$ or  $\hat{R}^{-1}{}_{23}$. With $\hat{R}$
the matrix  $\hat{R}^{-1}$ will satisfy the Yang-Baxter equation as well. Thus
we could have used $\hat{R}^{-1}$ in (\ref{qw2.8}) as well.
 
We now try to define a covariant transformation law:
\be
\omega(y_i)=S^l{}_i \otimes y_l
\label{qw2.19}
\ee
such that
\be
\omega(y_lx^l)=1  \otimes y_lx^l
\label{qw2.20}
\ee
This leads to
\be 
S=T^{-1}
\label{qw2.21}
\ee
Thus a covariant quantum plane exists if we can define $T^{-1}$.
Covariant relations for this quantum plane are of the form where $P$ 
is a projector
\be
y_a y_b P^{ba}_{lm}=0
\label{qw2.22}
\ee
Note the position of the indices.
\begin{eqnarray*}
&\omega(y_a y_b) P^{ba}{}_{lm}&={T^{-1}}^r{}_a {T^{-1}}^s{}_b \otimes y_r y_s
P^{ba}{}_{lm}\\
&&=P^{sr}{}_{ab}{T^{-1}}^b{}_m {T^{-1}}^a{}_l \otimes y_r y_s=0
\end{eqnarray*}
The second step follows from the RTT relations (\ref{qw1.4}).

It is possible to define covariant and consistent relations between the covariant
and contravariant quantum planes.
We start from the Ansatz:
\be 
x^r y_l=\Gamma^{mr}_{sl} y_m x^s
\label{qw2.23}
\ee

Covariance means that
\ba
&\omega(x^r y_l)&=T^r{}_s S^m{}_l \otimes x^s y_m=T^r{}_s S^m{}_l \otimes 
\Gamma^{ns}_{tm} y_n x^t \nonumber \\
&&= \Gamma^{mr}_{sl} S^n{}_m T^s{}_t \otimes y_n x^t
\label{qw2.24}
\ea
or:
\be
\begin{array}{rl}
T^r{}_a S^b{}_l \Gamma^{ca}_{db}=&\Gamma^{sr}_{tl} S^c{}_s T^t{}_d \\[2mm]
T^a{}_b  T^c{}_d  \Gamma^{bd}_{sr}=&\Gamma^{ac}_{db} T^d{}_s T^b{}_r
\end{array}
\label{qw2.25}
\ee

This is the condition for covariance. Now we have to prove consistency.

\ba
&0&=P^{ab}{}_{cd}x^c x^d y_l=P^{ab}{}_{cd}\Gamma^{ed}_{fl}x^c y_e x^f \nonumber \\
&&=P^{ab}{}_{cd} \Gamma^{ed}_{fl} \Gamma^{kc}_{me} y_k x^m x^f 
\label{qw2.26} \\
&&=\Gamma^{ka}_{\alpha \beta} \Gamma^{\beta c}_{\gamma l} 
P^{\alpha \gamma}{}_{mf} y_k x^m x^f \nonumber
\ea
For the last step we have to use the Yang-Baxter equation (\ref{qw2.14})
for $\Gamma$ and that $P$ is a polynomial in $\Gamma$.

Thus $\Gamma$ has to be a solution of the Yang-Baxter equation such that the 
$\Gamma TT$ relations hold for $T$s defined by the $\hat{R}TT$ relations (\ref{qw1.4}).
In general there might be several such solutions. Any such solution $\Gamma$ 
can be decomposed  into projectors on the invariant subspaces - this follows
 from covariance. We know that with $\hat{R}$, $\hat{R}^{-1}$ will always be
a solution as well.

If we use $\Gamma=q \hat{R}^{-1}$ we find that
$y_l x^l$ is central, it commutes with all $x^s$ and $y_r$.

\vskip 3mm

Another way to define a covariant transformation law is:
\be
\omega(\hat{y}_l)={\hat{S}}^k_l \otimes \hat{y}_k
\label{qw2.27}
\ee
and require that:
\be
\omega(x^l \hat{y}_l)=1 \otimes x^l \hat{y}_l
\label{qw2.28}
\ee
This leads to:
\be
\hat{S}=\widetilde{\tilde{T}^{-1}}
\label{qw2.29}
\ee
with the notation of (\ref{qw1.14}).

A similar analysis as above shows that

\be
\hat{y}_l x^k =\Gamma^{kr}_{ls} x^s \hat{y}_r
\label{qw2.30}
\ee
is consistent and covariant. $\Gamma$ again might be any solution of the 
Yang-Baxter equation that can be decomposed into covariant projectors. 
Compare the position of the matrices in (\ref{qw2.30}) and (\ref{qw2.23}). 
\be
\hat{y}_a \hat{y}_b P^{ba}{}_{lm}=0
\ee
is covariant again.

If we reorder $x$ and $y$ by (\ref{qw2.23}) we can relate (\ref{qw2.21}) and (\ref{qw2.29}).
If $\chi$ is the inverse matrix of $\Gamma$ 
\be
\chi^{ar}_{bs} \Gamma^{ms}_{lr}=\delta^a_l \delta^m_b
\label{qw2.32}
\ee
we find:
\be
y_a x^b=\chi^{br}_{al}x^ly_r
\label{qw2.33}
\ee
and, therefore
\be
y_a x^a=\chi^{ar}_{al}x^ly_r=x^l \hat{y}_l
\label{qw2.34}
\ee
Thus
\be
\hat{y}_l=\chi^{ar}_{al} y_r, \quad \quad \chi^{r}_{l}=\chi^{ar}_{al}
\label{qw2.35}
\ee
has to transform with $\hat{S}$ as in (\ref{qw2.27}).
A direct calculation for $SL_q(2)$ shows that this is consistent with (\ref{qw1.17}).
We have
\be
\widetilde{\tilde{T}^{-1}}=\chi^{-1}T^{-1}\chi
\label{qw2.36}
\ee
The change from (\ref{qw2.19}) to (\ref{qw2.27}) can be achieved by a linear change of the basis
in the quantum plane.

If, for $SL_q(2)$ we start from $\Gamma=\hat{R}$ in (\ref{qw2.23}) we find from (\ref{qw2.35}) that
in (\ref{qw2.30}) the respective $\Gamma$ is $\Gamma=\hat{R}^{-1}$. Thus by a direct computation
from
\be
x^r y_l = \hat{R}^{mr}_{sl} y_m x^s
\ee
it follows
\be
\hat{y}_a x^b=(\hat{R}^{-1})^{bc}_{ad} x^a \hat{y}_c
\ee

\section{Quantum derivatives}\label{sub2.3}

Another algebraic structure on co-modules is obtained by generalizing
the  Leibniz rule of derivatives:

\begin{equation}
\frac{\partial}{\partial x^i} x^j = \delta^j_i + x^j 
\frac{\partial}{\partial x^i}\label{qw1}
\end{equation}
We demand that the algebra 
generated by the elements of the quantum plane algebra $x^i$ and the derivatives $\partial_i$, divided by proper ideals,
has the Poincar\'{e}-Birkhoff-Witt property.In addition
we shall show that there is an exterior differential calculus based on these quantum derivatives.

We start with an Ansatz:
\begin{equation}
\partial_ix^j = \delta^j_i + C^{jk}_{il} x_k\partial^l\label{qw2}
\end{equation}
It allows arbitrary coefficients $C^{jk}_{il}$ that should take care of the noncommutativity
of the space. Covariance and the PBW requirement will determine the $n^2 \times n^2$ matrix $C$ to a large extent.

Covariance will be achieved by requiring $\partial_i$ to transform covariantly:
\begin{equation}
\omega (\partial_i) = \hat{S}^l{}_i \otimes\partial_l   .
\end{equation}
This leads to certain conditions on $C$ which we derive now:
\begin{eqnarray*}
\omega(\partial_ix^j) &=& \hat{S}^l{}_iT^j{}_k \otimes\partial_lx^k\\
&=& \hat{S}^l{}_iT^j{}_k \otimes\{\delta^k_l + C^{km}_{ln} x^n \partial_m\}
\end{eqnarray*}
\[
\omega(\delta^j_i + C^{jk}_{il} x^l\partial_k) = \delta^j_i + C^{jk}_{il} T^l{}_n{S}^m{}_k \otimes x^n\partial_m
\]
Equating these expressions yields:
\begin{equation}
\hat{S}^l{}_iT^j{}_kC^{km}_{ln} = C^{jk}_{il}T^l{}_n\hat{S}^m{}_k \quad
\makebox{and}\quad \hat{S}^j{}_aT^b{}_j = \delta_a{}^b ,\quad  T^j{}_a\hat{S}^b{}_j = \delta_a{}^b
\label{qw3.4}
\end{equation}
Thus $\hat{S}$ is as in (\ref{qw2.27}). As  a further consequence of (\ref{qw3.4}):
\begin{equation}
C^{ab}_{cd}T^c{}_r T^d{}_s = T^a{}_c T^b{}_d C^{cd}_{rs}\label{qw5}
\end{equation}

Eqn. (\ref{qw5}) can be satisfied if $C$ is a polynomial in $\hat{R}$.
To guarantee covariance such a polynomial  has to be a
combination of projectors:
\begin{equation}
C = \sum_{\mbox{all projectors}} c_lP_l\label{qw6}   
\end{equation}

Next we demand consistency with the algebraic relations for the quantum plane:
\begin{equation}
P_A{}^{ab}{}_{cd}x^cx^d = 0\label{qw7}
\end{equation}
$P_A$ is the antisymmetrizer. We compute
\begin{equation}
\partial_i P_A{}^{ab}{}_{cd}x^cx^d =\label{qw8}
\end{equation}
\[
( P_A{}^{ab}{}_{ij} + P_A{}^{ab}{}_{cd}  C^{cd}{}_{ij}) x^j
+  P_A{}^{ab}{}_{cd} C^{cr}_{im} C^{ds}_{rn} x^mx^n\partial_s
\]

Consistency requires that this expression should be zero. There are two equations to be satisfied. The first one is:
\begin{equation}
P_A + P_AC = 0\label{qw9}
\end{equation}
{From} (\ref{qw9}) follows:
\begin{equation}
C = - P_A +  \sum_{\mbox{symmetric projectors}} c_lP_l\label{qw10}
\end{equation}

The second equation that we obtain is more difficult to analyse. We realize that the last term in eqn (\ref{qw8})
 can be written as follows
\[
(P_{A12} C_{23} C_{12})^{\alpha_1\alpha_2\alpha_3}{}_{\beta_1\beta_2\beta_3} x^{\beta_2}x^{\beta_3}
\partial_{\alpha_3}
\]
Here we use the notation of (\ref{qw2.12}), (\ref{qw2.13}). If we manage to carry $P_A$ to the right hand side as a $P_{A23}$, then $P_A$ would act
on: $x^{\beta_2}x^{\beta_3}$ and give zero. This hints at the structure of a Yang-Baxter equation:

\begin{equation}
P_{A12} C_{23} C_{12} = C_{23} C_{12} P_{A23}
\end{equation}
If $C$ is a solution of the Yang-Baxter equation:
\begin{equation}
C_{12} C_{23} C_{12} = C_{23} C_{12} C_{23}
\end{equation} 
then it would be true that for any polynomial ${\cal P}(C)$ we would have:
\begin{equation}
{\cal P}(C)_{12} C_{23} C_{12} = C_{23} C_{12} {\cal P}(C)_{23}.
\end{equation} 
We conclude that if $C$ has the structure given in eqn(\ref{qw10}) and if coefficients $c_l$ can be found such that $C$ 
satisfies the Yang-Baxter equation and if all the $c_l \not= - 1$ ( so that we can write the projector $P_A$ as a 
polynomial in $C$)  then the derivative defined in (\ref{qw2}) is consistent with the quantum plane relations (\ref{qw7}). This is
the case for $SL_q(n)$. There we have eqn (\ref{qw10}):
\[
\hat{R} = qS - \frac{1}{q} A ,\quad  \hat{R}^{-1} =  \frac{1}{q} S - qA
\]
and therefore:
\begin{equation}
C = q\hat{R}\quad \mbox{or} \quad C^{-1} = \frac{1}{q} \hat{R}^{-1}.
\end{equation}
There are two solutions that have the desired properties.

We conclude that
\begin{equation}
\partial_ix^j = \delta^j_i + q\hat{R}^{jk}{}_{il} x^l\partial_k\label{qw15}
\end{equation}
or
\begin{equation}
\hat{\partial}_ix^j = \delta^j_i + \frac{1}{q}\hat{R}^{-1jk}{}_{il} x^l\hat{\partial}_k
\end{equation}
are two possibilities to define a covariant derivative on a quantum plane consistent with the defining
relations of $SL_q(n)$ quantum planes.

In general, if there are consistent and covariant derivatives, i.e. if we can find a matrix $C^{ab}{}_{cd}$ that satisfies
the Yang-Baxter equation and that has the decomposition in projectors (\ref{qw10}), then
\begin{equation}
C^{-1} = - P_A +  \sum_{\mbox{sym.}} c^{-1}_l P_l
\end{equation}
has the same properties and we have  two possibilities to define covariant derivatives.

To complete the algebra we have to find the $\partial\partial$ relations. To this end we compute:
\[
\partial_j\partial_ix^ax^b = \delta^a_i \delta^b_j + C^{ab}_{ij} + ...
\]
The unit $\delta^a_i \delta^b_j$ has a projector decomposition:
\[
1 = P_A +  \sum_{\mbox{sym.}} P_l
\]
Combining this with (\ref{qw10}) we find
\[
\partial_j\partial_ix^ax^b = \sum_{\mbox{sym.}} (c_l+1) P_l{}^{ab}{}_{ij} + ...
\]
and we conclude that it would be  consistent  to demand
\begin{equation}\label{qw3.18}
\partial_j \partial_i P_A{}^{ij}{}_{rs} = 0.
\end{equation}
This is a covariant condition and we can show by an  argument similar to the one 
that followed eqn (\ref{qw8})  that (\ref{qw3.18})
 is consistent with (\ref{qw15}),  including all the terms that have been indicated by dots. 

We summarize
\begin{equation}
P_A{}^{ab}{}_{cd} x^cx^d = 0
\end{equation}
\[
\partial_a \partial_b P_A{}^{ba}{}_{cd}  = 0
\]
\[\partial_a x^b = \delta^b_a + C^{bc}_{ad} x^d \partial_c
\]
and
\begin{equation}
C = -P_A +   \sum_{\mbox{sym.}} c_lP_l
\end{equation}
\[ 
C_{12}C_{23} C_{12} = C_{23} C_{12}C_{23}
\]
It defines a
covariant and consistent algebra. The matrix $C$ can be replaced by $C^{-1}$ and we again obtain a covariant
and consistent algebra.
\begin{equation}
\hat{\partial}_a \hat{\partial}_b P_A{}^{ba}{}_{cd} = 0
\end{equation}
\[
\hat{\partial}_a  x^b = \delta^b_a + C^{-1bc}{}_{ad} x^d\hat{\partial}_c
\]

We can consider the algebra generated by  $x,\partial$ and $\hat{\partial}$.
 A similar argument that led to (\ref{qw3.18}) shows
that 
\begin{equation}
\hat{\partial}_a \partial_b = C^{cd}{}_{ba} \partial_d\hat{\partial}_c
\label{qw22}
\end{equation}
is a consistent and covariant condition.

Let me list the relations for the algebra based on the $\hat{R}$ matrix (\ref{qw5}), i.e. for $SL_q(2)$:
\begin{eqnarray}
x^ix^j &=& \frac{1}{q} \hat{R}^{ij}{}_{kl} x^kx^l:  
\label{qw23} \\
x^1x^2 &=& qx^2x^1\nn
\rule{0mm}{8mm}
\partial_i x^j &=& \delta^j_i + q\hat{R}^{jk}_{il} x^l\partial_k  : \label{qw24}\\
\partial_1x^1 &=& 1 + q^2x^1\partial_1 + q\lambda x^2\partial_2\nn
\partial_1x^2 &=&  qx^2\partial_1\nn
\partial_2x^1 &=&  qx^1\partial_2\nn
\partial_2x^2 &=& 1 + q^2x^2\partial_2\nn
\rule{0mm}{8mm}
\partial_a\partial_b &=& \frac{1}{q} \partial_c\partial_d \hat{R}^{dc}{}_{ba}   : \label{qw25}\\
\partial_1\partial_2 &=& \frac{1}{q} \partial_2\partial_1\nn
\rule{0mm}{8mm}
\hat{\partial}_i x^j &=& \delta^j_i +  \frac{1}{q} \hat{R}^{-1jk}{}_{il} x^k \partial_l  : \label{qw26}\\ 
\hat{\partial}_1 x^1 &=& 1 +  \frac{1}{q^2} x^1 \hat{\partial}_1\nn
\hat{\partial}_1 x^2 &=&  \frac{1}{q} x^2 \hat{\partial}_1\nn
\hat{\partial}_2 x^1 &=&  \frac{1}{q} x^1 \hat{\partial}_2\nn
\hat{\partial}_2 x^2 &=& 1 +  \frac{1}{q^2} x^2 \hat{\partial}_2 - \frac{\lambda}{q} x^1\hat{\partial}_1\nn
\rule{0mm}{8mm}
\hat{\partial}_a\hat{\partial}_b &=& \frac{1}{q} \hat{\partial}_c\hat{\partial}_d \hat{R}^{dc}{}_{ba}   : \label{qw27}\\
\hat{\partial}_1\hat{\partial}_2 &=& \frac{1}{q}\hat{\partial}_2\hat{\partial}_1\nn
\rule{0mm}{8mm}
\hat{\partial}_a\partial_b &=& q \hat{R}^{cd}{}_{ba} \partial_d\hat{\partial}_c   :\nn
\hat{\partial}_1\partial_1 &=& q^2 \partial_1\hat{\partial}_1  \nn
\hat{\partial}_1\partial_2 &=& q \partial_2\hat{\partial}_1\nn
\hat{\partial}_2\partial_1 &=& q \partial_1\hat{\partial}_ 2+ \lambda q \partial_2\hat{\partial}_1\nn
\hat{\partial}_2\partial_2 &=& q^2 \partial_2\hat{\partial}_2\nonumber
\end{eqnarray}

There is an exterior differential calculus based on these quantum derivatives . We introduce differentials and as a generalization of the anticommutativity of ordinary differentials we demand:
\begin{equation}
P_S{}^{ab}{}_{cd} dx^cdx^d = 0\label{qw28}
\end{equation}
and as usual:
\begin{equation}
d = dx^i\partial_i\label{qw29}
\end{equation}
\be
d^2 = 0  ,\quad   d(fg) = (df)g + fdg  ,\quad  ddx^i = - dx^i d\label{qw30}
\ee
We can make use of (\ref{qw28}) and (\ref{qw29}) to find relations for $x$ and $dx$. We start with an Ansatz:
\begin{equation}
x^idx^j = O^{ij}_{ks}dx^kx^s
\end{equation}
Acting with $d$ on this equation yields:
\[
dx^idx^j = -  O^{ij}_{ks}dx^k dx^s
\]
We combine this with (\ref{qw28}) and obtain
\[
(1+O) =  \sum_{\mbox{sym.}} c_lP_l
\]
and in turn:
\begin{equation}
O = - P_A + \sum_{\mbox{sym.}} (c_l -1)P_l\label{qw32}
\end{equation}
Next we evaluate the equation:

\begin{eqnarray}
d \cdot x^i &=& dx^i + x^id\nn
dx^l\partial_lx^i &=& dx^i + x^idx^l\partial_l\nn
dx^lC^{ir}{}_{ls} x^s\partial_r &=& O^{ir}{}_{ls} dx^lx^s\partial_r\nonumber
\end{eqnarray}
We conclude:
\begin{equation}
O^{ir}{}_{ls} = C^{ir}{}_{ls} 
\end{equation}
This is consistent with (\ref{qw32}).

We could also have derived the exterior algebra:
\begin{eqnarray}
P_A{}^{ij}{}_{kl} x^kx^l &=& 0\\
P_S{}^{ij}{}_{kl} dx^kdx^l &=& 0\nonumber
\end{eqnarray}
\[
x^idx^j = C^{ij}{}_{kl} dx^kx^l
\]
from consistency arguments and the properties of the differential as given by (\ref{qw30}).
The derivatives and their properties then would have followed from (\ref{qw29}). For our purpose,
 however, the existence
of derivatives and their properties is more essential. That there is an exterior calculus with all the
 properties (\ref{qw30}) -
especially the unchanged Leibniz rule - is a pleasant surprise. 

We can now deal with the entire algebra generated
by $x^i,dx^j, \partial_l$ and divided by the respective ideals. For this purpose the $dx^l, \partial_j$ relations have to 
be specified.

We again start with an Ansatz:
\[
\partial_jdx^i = D^{ik}{}_{jl} dx^l\partial_k
\]
and we multiply this equation by $x^r$ from the right:
\[
\partial_jdx^ix^r = D^{ik}{}_{jl} dx^l\partial_kx^r
\]
\[
C^{-1ir}{}_{st} \partial_jx^sdx^t = D^{ik}{}_{jl}dx^l\partial_kx^r
\]
This equation splits into two parts:
\[
C^{-1ir}{}_{jt} dx^t = D^{ir}{}_{jt}dx^t
\]
and
\[
C^{-1ir}{}_{st}  C^{sa}{}_{jb} x^b\partial_adx^t = D^{ik}{}_{jl} C^{la}{}_{kb} dx^lx^b\partial_a
\]
>From the first equation we conclude:
\begin{equation}
D^{ir}{}_{jt} = C^{-1ir}{}_{jt} 
\end{equation}
The second equation is true because $C$ satisfies the Yang-Baxter equation. For the $x,\partial$ 
and $dx$ algebra to be 
consistent it is now necessary that $C$ satisfies the Yang-Baxter equation. The result is:
\begin{equation}
\partial_j dx^i =C^{-1ik}{}_{jl} dx^l\partial_k
\end{equation}
It should be noted that whereas $x$ and $d$ have simple commutation properties $(dx^a = (dx^a) + x^ad)$, 
this is not
true for $d$ and $\partial_i$.

We compute:
\begin{equation}
\partial_id = \partial_idx^l\partial_l = C^{-1la}{}_{ib} dx^b\partial_a\partial_l\label{qw37} 
\end{equation}
The antisymmetric projector of $C$ does not contribute to (\ref{qw37}). The result, however, depends on
$C$ via  the symmetric projectors. For $SL_q(n)$ it can be evaluated explicitely:
\[
C^{-1} = q^{-1}\hat{R}^{-1} = -A + q^{-2}S = q^{-2}1 - (1+q^{-2})A
\]
and we obtain for $SL_q(n)$:
\begin{equation}
\partial_id = q^{-2}d\partial_i
\end{equation}
To complete the list of explicit relations for  $SL_2(2)$ we finish this chapter by giving the $xdx$ relations:
\begin{eqnarray}
dx^1x^1 &=& q^2x^1dx^1\nn
dx^1x^2 &=& qx^2dx^1 + (q^2-1) x^1dx^2\\
dx^2x^1 &=& qx^1dx^2\nn
dx^2x^2 &=& q^2x^2dx^2\nonumber
\end{eqnarray}

\section{Conjugation}\label{sub2.4}

Let us finally enrich the algebraic structure by adding a conjugation.
For a physical interpretation this will be essential because such an
interpretation will rest on Hilbert space representations of the algebra 
and observables will have to be identified with essentially self-adjoint
operators in Hilbert space. The conjugation defined here is a purely algebraic
operation to start with but later has to be identified with mapping
 to the adjoint
operator in Hilbert space.  We 
introduce conjugate variables as new independent elements of the algebra:
\begin{equation}
\overline{x^i} \equiv \overline{x}_i\quad , \quad
\overline{x^i x^j} = \overline{x}_j \overline{x}_i\label{qw4.1}
\end{equation}
The lower index of $\overline{x}_i$ does not mean that they
 transform covariantly as defined by (\ref{qw2.19}) or (\ref{qw2.27}).
The transformation law follows from (\ref{qw2.1}):
\begin{equation}
\omega(\overline{x}_i) = \overline{T^i{}_k} \otimes \overline{x}_k =
\overline{T}^k{}_i \otimes \overline{x}_k\label{qw4.2}
\end{equation}
We have defined:
\begin{equation}
\overline{T^i{}_k} \equiv  \overline{T}^k{}_i\label{qw4.3}
\end{equation}
{From} the $\hat{R}TT$ relations (\ref{qw1.4}) follows by conjugating:
\begin{equation}
\hat{R}^{+sr}{}_{lk} \overline{T}^l{}_j \overline{T}^k{}_i = \overline{T}^s{}_l \overline{T}^r{}_k \hat{R}^{+lk}{}_{ji}\label{qw4.4}
\end{equation}
$\hat{R}^{+sr}$ is defined as follows:
\begin{equation}
\hat{R}^{+sr}{}_{lk} = \overline{\hat{R}^{kl}{}_{rs}}\label{qw4.5}
\end{equation}
$\hat{R}^+$ satisfies the Quantum Yang-Baxter equation if $\hat{R}$ does.
This can be verified by direct calculation.

For the $\hat{R}$ matrix (\ref{qw1.5}), (\ref{qw1.12}) and real $q(\overline{q} = q)$ we find:
\begin{equation}
\hat{R}^{+sr}{}_{lk} =  \hat{R}^{rs}{}_{lk}\label{qw4.6}
\end{equation}
$\hat{R}^+$ in this case is obtained from $\hat{R}$ by a symmilarity transformation and has the same eigenvalues.

The $\overline{x}$ $\overline{x}$ relations are obtained by conjugating 
(\ref{qw2.6}):
\begin{equation}
\overline{x}_j\overline{x}_i = \frac{1}{q}  \hat{R}^{kl}{}_{ij}\overline{x}_l\overline{x}_k
\label{qw4.7}
\end{equation}

This is exactly of the same type as the relation (\ref{qw2.22}) with
 $P = A$ of (\ref{qw1.10}).
A consistent $x \overline{x}$ relation is therefore obtained from (\ref{qw2.23})
 by
$\Gamma = q \hat{R}^{-1}$.
We choose this possibility to have $\overline{x}_i x^i$ central:
\begin{equation}
x^i\overline{x}_j = q \hat{R}^{-1li}{}_{kj} \overline{x}_lx^k\label{qw4.8}
\end{equation}
This is consistent with conjugation as well.

We demand covariance of this relation with respect to (\ref{qw2.1}) 
and (\ref{qw4.2}).
\begin{equation}
\omega(x^i \overline{x}_j) = T^i{}_l \overline{T}^k{}_j x^l \overline{x}_k =
\label{qw4.9}
\end{equation}
\[
q T^i{}_l \overline{T}^k{}_j (\hat{R}^{-1})^{rl}{}_{sk} \overline{x}_rx^s
\]
\[
q (\hat{R}^{-1})^{li}{}_{kj} \overline{T}^r{}_l T^k{}_s  \overline{x}_rx^s
\]
This leads to:
\begin{equation}
T^i{}_l \overline{T}^k{}_j (\hat{R}^{-1})^{rl}{}_{sk} = 
(\hat{R}^{-1})^{li}{}_{kj} \overline{T}^r{}_l T^k{}_s\label{qw4.10}  
\end{equation}

These are the $\hat{R}T\overline{T}$ relations. All this could be put together to one big quantum plane of $2n$ elements
$(x^i, \overline{x}_j)$ and one big $(2n)^2 \times (2n)^2 \hat{R}$ matrix 
satisfying the corresponding Yang-Baxter
equation. What we arrive at is the complex quantum plane generated by $x^i, \overline{x}_j$ and divided by the
ideals generated by the relations (\ref{qw2.6}), (\ref{qw4.7}) and (\ref{qw4.8}).
 The quantum group represented by $T$ and
 $\overline{T}$ is $GL_q(2,C)$ or, for
$det_q T = 1$, $SL_q(2,C)$.

A generalization to several quantum planes is possible starting from
 (\ref{qw2.8}).
\begin{equation}
x^iy^j = \frac{\kappa}{q} \hat{R}^{ij}{}_{kl} y^kx^l\label{qw4.11}
\end{equation}
\[
\overline{y}_j \bar{x}_i =  \frac{\kappa}{q} \hat{R}^{kl}{}_{ij} \overline{x}_l \bar{y}_k
\]
\[ 
x^i\bar{y}_j = \kappa q {\hat{R}^{-1}}{}^{li}_{kj} \bar{y}_lx^k
\]
\[
y^j\bar{x}_i = \kappa q {\hat{R}^{-1}}{}^{kj}_{li} \bar{x}_ky^l
\]
For $n=2$ the explicit $x\bar{y}$ relations are (for $\kappa = 1$):
\begin{equation}
x^1\bar{y}_1 = \bar{y}_1x^1 - q\lambda \bar{y}_2x^2\label{qw4.12}
\end{equation}
\[
x^1\bar{y}_2 =  q \bar{y}_2x^1
\]
\[
x^2\bar{y}_1 =  q \bar{y}_1x^2
\]
\[
x^2\bar{y}_2 =  \bar{y}_2x^2
\]

The $\bar{y} \bar{x}$ relations follow from (\ref{qw2.9}) by conjugating.

For the entries of $T, \bar{T}$, as they are defined by (\ref{qw1.1})
 and its conjugate the  $T, \bar{T}$ relations are:
\begin{equation}
a \bar{a} = \bar{a} a - q\lambda \bar{c} c\label{qw4.13}
\end{equation}
\[
a\bar{b} = \frac{1}{q}\bar{b}a - \lambda \bar{d} c
\]
\[
a\bar{c} = q\bar{c}a
\]
\[
a\bar{d} = \bar{d}a
\]

\[
b\bar{a} = \frac{1}{q} \bar{a}b - \lambda \bar{c} d
\]
\[
b\bar{b} = \bar{b}b + q\lambda (\bar{a}a - \bar{d}d - q\lambda \bar{c} c)
\]
\[
b\bar{c} = \bar{c}b
\]
\[
b\bar{d} = q\bar{d} b + \lambda q^2 \bar{c} a
\]
\[
c\bar{a} = q\bar{a} c
\]
\[
c\bar{b} = \bar{b} c
\]
\[
c\bar{c} = \bar{c} c
\]
\[
c\bar{d} = \frac{1}{q} \bar{d}c
\]

\[
d\bar{a} = \bar{a} d
\]
\[
d\bar{b} = q\bar{b} d + \lambda q^2 \bar{c} a
\]
\[
d\bar{c} = \frac{1}{q} \bar{c}d
\]
\[
d\bar{d} = \bar{d} d + \lambda q \bar{c} c
\]

A comparison of (\ref{qw4.2}) with (\ref{qw2.19}) shows
 that it is possible to identify 
$\overline{T}$ with $T^{-1}$. This then defines the quantum group
$U_q(n)$ or for $det_q T=1$ the quantum group $SU_q(n)$.

For $n = 2$,

\be
T=\left(\begin{array}{cc} a & b\\ c & d\end{array} \right) \quad ,
\quad T^{-1}=\left(\begin{array}{cc} d & {-1 \over q}b \\ -qc & a\label{qw4.14}
\end{array} \right)
\ee

we find

\begin{eqnarray}
&\overline{a}=d,  & \overline{b}=-qc \nonumber \\
&\overline{d}=a,  & \overline{c}=-{1 \over q} b\label{qw4.15}
\end{eqnarray}

It can be verfied directly that (\ref{qw4.15}) is consistent with (\ref{qw4.13}).

To extend conjugation to an algebra with derivatives we start with
differentials. From (\ref{qw28}) we have
\be
x^c dx^d = q \hat{R}^{cd}{}_{ab}dx^a x^b\label{qw4.16}
\ee
and therefore
\be
dx^c dx^d = - q \hat{R}^{cd}{}_{ab}dx^a dx^b\label{qw4.17}
\ee
Conjugation leads to
\be
\overline{dx}_d \overline{x}_c = q \hat{R}^{ab}{}_{cd}\overline{x}_b
\overline{dx}_a\label{qw4.18}
\ee 
and
\be
\overline{dx}_d \overline{x}_d = -q \hat{R}^{ab}{}_{cd}\overline{dx}_b
\overline{dx}_a\label{qw4.19}
\ee 
{From} (\ref{qw4.11}) we conclude:
\be
x^i \overline{dx}_j = q {\hat{R}^{-1}}{}^{li}_{kj} \overline{dx}_l 
 {x}^k\label{qw4.20}
\ee 
and
\be
dx^i \overline{dx}_j = -q {\hat{R}^{-1}}{}^{li}_{kj} \overline{dx}_l 
{dx}^k\label{qw4.21}
\ee 

\section{$q$-Deformed Heisenberg algebra}\label{sub2.5}

The canonical commutation relations are at the basis of a  quantum 
mechanical system:
\be
[\hat{x}, \hat{p}] =\hat{x} \hat{p} -\hat{p} \hat{x} = i\label{qw5.1}
\ee
The elements of this algebra are supposed to be selfadjoint
\be
\overline{\hat{x}} = \hat{x} \quad , \quad \label{qw5.2}
\overline{\hat{p}} = \hat{p}
\ee
A physical system is defined through a representation of this algebra 
in a Hilbert space where selfadjoint elements of the algebra have to be
represented by (essentially) selfadjoint linear operators.

In quantum mechanics the elements of (\ref{qw5.1}) are represented in the Hilbert space of square-integrable functions by:
\be
\hat{x} = x \quad ,\quad  \hat{p}= - i \frac{\partial}{\partial x}\label{qw5.3}
\ee
Starting from the algebra (\ref{qw5.1}), the spectrum of the linear operator
$\hat{x}$ can be interpreted  as the manifold on which the 
physical system lives -
i.e. the configuration space.

In quantum mechanics it is $\mathbb{R}_1 (x \in \mathbb{R}_1)$. The
element $\hat{p}$ is a differential operator on this manifold.

We shall change the algebra (\ref{qw5.1}) in accord with quantum group 
considerations. It is natural to assume that $\hat{x}$ is an element of a
quantum plane and to relate $\hat{p}$ to a derivative in such a plane. The
simplest example that we can consider is suggested by the last equation of 
(\ref{qw25}). With an obvious change in notation we study the algebra
\be
\partial x = 1 + q x \partial\label{qw5.4}
\ee
More precisely, we study the free algebra generated by the elements $x$ and 
$\partial$ and divided by the ideal generated by (\ref{qw5.4}).

If we assume $x$ to be selfadjoint
\be
\overline{x} = x
\label{qw5.5}
\ee
we see that this cannot be the case for $i\partial$ because we find from 
(\ref{qw5.4}) that:
\be
\overline{\partial} x =- \frac{1}{q} + \frac{1}{q} x 
\overline{\partial}\label{qw5.6}
\ee
In general $\overline{\partial}$ will be related to $\hat{\partial}$ rather
than to $\partial$. (See the second equation of (\ref{qw26})). Thus we could
study the algebra generated by $x$, $\partial$ and $\overline{\partial}$ and
divide by (\ref{qw5.4}), (\ref{qw5.6}) and an ideal generated by 
$\partial \overline{\partial}$ relations. These can be found by a similar
argument that led to the $\partial \hat{\partial}$ relations 
(\ref{qw22}). We compute from (\ref{qw5.4}) and (\ref{qw5.6})  
$\partial \overline{\partial}x$ and $\overline{\partial} \partial x$ and
find that
\be
\overline{\partial} \partial = q \partial \overline{\partial}\label{qw5.7}
\ee
is consistent with these calculations. If we now try to define
an operator $\hat{p}$ by $\hat{p}=-\frac{i}{2}(\partial -\overline{\partial})$
we find that the $x, \hat{p}$ relations do not 
close. The real part of $\partial$ has to be introduced as well. Thus our
Heisenberg algebra would have one space and two momentum operators - a system
that will hardly find a physical interpretation.

It turns out that  $\overline{\partial}$ can be related to $\partial$ and $x$
in a nonlinear way. This relation involves the scaling operator 
$\Lambda$:
\ba
\Lambda &\equiv &q^{\frac{1}{2}} \left( 1+(q-1) x \partial \right) \nonumber \\
\Lambda x&=&qx \Lambda
\label{qw5.8} \\
\Lambda \partial &=&q^{-1} \partial \Lambda \nonumber
\ea
The scaling property follows from (\ref{qw5.4}). 

We now define
\be
\tilde{\partial}=-q^{-\frac{1}{2}} \Lambda^{-1} \partial
\label{qw5.9}
\ee
$\Lambda^{-1}$ is defined by an expansion in $(q-1)$. We find
\be
\begin{array}{rcl}
\tilde{\partial} x&=&-{1 \over q} +{1 \over q} x \tilde \partial \\[2mm]
\tilde{\partial} \partial&=&q \partial \tilde{\partial}
\end{array}
\label{qw5.10}
\ee

Comparing this with (\ref{qw5.6}) and (\ref{qw5.7}) it follows from 
(\ref{qw5.9}) and (\ref{qw5.10}) that conjugation in the $x, \partial$ algebra can
be defined by
\be
\overline{x}=x \quad ,  \quad \overline{\partial}=-q^{-\frac{1}{2}} \Lambda^{-1} \partial
\label{qw5.11}
\ee

Conjugating $\Lambda$ and using (\ref{qw5.11}) shows that 
\be
\overline{\Lambda} = \Lambda^{-1}\label{qw5.12}
\ee
$\Lambda$ is a unitary element of the algebra, 
this justifies the factor $q^{\frac{1}{2}}$ in the
definition of $\Lambda$.

The existence of a scaling operator $\Lambda$ 
and the definition of the conjugation 
(\ref{qw5.11}) seems to be very specific for the $x, \partial$ algebra    
(\ref{qw5.4}). It is however generic in the sense that a scaling operator and a definition of conjugation based on it can be found for all the quantum planes
defined by $SO_q(n)$ and $SO_q(1,n)$.

The definition of the $q$-deformed Heisenberg algebra will now be based on the
definition of the momentum:
\be
p = - \frac{i}{2} (\partial - \overline{\partial})\label{qw5.13}
\ee
It is selfadjoint. From the $x, \partial$ algebra and the definition of 
$\overline{\partial}$ follows
\be
\begin{array}{ll}
q^{\frac{1}{2}} xp -q^{-\frac{1}{2}} px &=i \Lambda^{-1} \\[2mm]
\Lambda x=qx\Lambda  & \quad \Lambda p=q^{-1} p \Lambda
\end{array}
\label{qw5.14}
\ee
and
\be
\overline{p} = p \quad , \quad \overline{x} = x  \quad , \quad
\overline{\Lambda} = \Lambda^{-1}\label{qw5.15}
\ee

These algebraic relations can be verified in the $x, \partial$ representation
where the ordered $x, \partial$ monomials form a basis. We shall take
(\ref{qw5.14}) and (\ref{qw5.15}) as the defining relations for the $q$-deformed
Heisenbergalgebra without making further reference to its 
$x, \partial$ representation.

\section{The $\mathbf{q}$-deformed Lie algebra $\mathbf{sl_q(2)}$}\label{sub2.6}

The algebra $sl_q(2)$ is the dual object to the quantum group $SL_q(2)$.
In this chapter we first define this algebra without referring to the quantum
group and then study its representations.

The algebra is defined as follows:
\ba
{1 \over q} T^+ T^- -q T^- T^+ &=&T^3 \nonumber \\
q^2 T^3 T^+ - {1 \over q^2} T^+ T^3&=&(q+{1 \over q}) T^+
\label{qw7.1} \\
{1 \over q^2} T^3 T^- -q^2 T^- T^3 &=&-(q+{1 \over q}) T^- \nonumber
\ea
It is often convenient to introduce the ``group like'' element $\tau$:
\ba
&\tau=1-\lambda T^3, \quad & \lambda=q-{1 \over q}
\label{qw7.2} \\
&T^3 =\lambda^{-1}(1-\tau) \nonumber
\ea
{From} (\ref{qw7.1}) follows:
\ba
\tau T^+&=&{1 \over q^4} T^+ \tau \nonumber \\
\tau T^-&=&q^4 T^- \label{qw7.3} \\
{1 \over q} T^+ T^- -q  T^- T^+&=&{1 \over \lambda} (\tau -1) \nonumber
\ea
The algebra (\ref{qw7.1}) has a Casimir operator
\be
\vec{T}^2=q^2 (T^- T^+ +{1 \over {\lambda^2}}) \tau^{-1/2}+{1 \over {\lambda^2}} (\tau^{1/2}-1-q^2)
\label{qw7.4}
\ee
$\vec{T}^2$ commutes with $T^+, T^-$ and $T^3$.
The constant term has been added to give the usual Casimir operator in the limit $q \rightarrow 1$.

The algebra (\ref{qw7.1}) allows a coproduct:
\be
\begin{array}{rcl}
\Delta(T^3) &=&T^3 \otimes 1 + \tau \otimes T^3 \\[1mm]
\Delta(T^{\pm})&=&T^{\pm} \otimes 1 + \tau^{1/2} \otimes T^{\pm} \nonumber
\end{array}
\label{qw7.5}
\ee
This follows from its construction as the dual object of the Hopf algebra
$SL_q(2)$ or it can be verified directly that:
\ba
\hq \Delta(T^+) \Delta(T^-)-q\Delta(T^-)\Delta(T^+)&=&\Delta(T^3) \nn
q^2 \Delta(T^3) \Delta(T^+)-\frac{1}{q^2} \Delta(T^+)\Delta(T^3)&=&(q+\hq) \Delta(T^+) 
\label{qw7.6} \\
\frac{1}{q^2} \Delta(T^3) \Delta(T^-) -q^2 \Delta(T^-)\Delta(T^3)&=&-(q+\hq) \Delta(T^-) \nonumber
\ea
To complete the definition of $sl_q(2)$ as a Hopf algebra we add the definition
of the co-unit:
\be
\varepsilon(T)=0
\ee
and the antipode:
\be
\begin{array}{rcl}
S(T^{\pm})&=&-T^{\pm} \tau^{-1} \\[1mm]
S(T^3)&=&-T^3 \tau^{-1} 
\end{array}
\ee

For the Hopf algebra $su_q(2)$ conjugation properties have to be assigned. It
is easy to see that
\be
\overline{T}^3=T^3, \quad \overline{T}^+=\frac{1}{q^2} T^-, \quad 
\overline{T}^-=q^2 T^+
\label{qw7.9}
\ee
is compatible with all previous relations.

With the very same methods with which the representations of angular momentum
are constructed in quantum mechanics we can construct representations of 
$su_q(2)$. The representations are characterized by the eigenvalue of the Casimir operator and the states in a representation by the eigenvalues of $T^3$.

We first define the $q$ number
\be
[n]={q^n-q^{-n} \over q-q^{-1}}
\label{qw7.10}
\ee
and list the non-vanishing matrix elements:
\ba
\vec{T}^2 | j,m > &=& q [j][j+1] |j,m> \nonumber \\
T^3  | j,m > &=& q^{-2m} [2m]   | j,m > \nonumber \\
T^+  | j,m > &=& q^{-m-\frac{3}{2}} \sqrt{[j+m+1][j-m]} | j,m+1>
\label{qw7.11} \\
T^-  | j,m > &=& q^{-m+\frac{3}{2}} \sqrt{[j+m][j-m+1]} | j,m-1> \nonumber \\
\tau | j,m > &=& q^{-4m}  | j,m > \nonumber
\ea

The representation, characterized by $j$ is $2j+1$ dimensional and it is easy to see that in the limit $q \rightarrow 1$ it becomes the representation of the
usual angular momentum $j^i$:
\be
q \rightarrow 1: T^3 \rightarrow 2 j_3, \quad T^{\pm} \rightarrow j^{\pm}
\label{qw7.12}
\ee
such that
\ba
j^+ j^- -j^- j^+&=&2 j_3 \nn
j_3 j^+-j^+j_3&=&j^+
\label{qw7.13} \\
j^- j_3-j_3j^-&=&j^- \nonumber
\ea
It is remarkable that the non-vanishing matrix elements are exactly the same
as for the representations of the undeformed angular momentum. This suggests
 that the $T$ matrices are in the enveloping algebra of the $j$ matrices. For
the representations of the algebra (\ref{qw7.13}) we know that
\ba
j_3  | j,m > &=& m | j,m > \nonumber \\
j^+  | j,m > &=& \sqrt{(j+m+1)(j-m)} | j,m+1>
\label{qw7.14} \\
j^-  | j,m > &=& \sqrt{(j+m)(j-m+1)} | j,m-1> \nonumber
\ea
or
\be
\begin{array}{c}
 | j,m+1> = {\displaystyle 1 \over \displaystyle \sqrt{j(j+1)-m^2-m}} j^+  | j,m >
={\displaystyle 1 \over \displaystyle \sqrt{\vec{j}^2 -j_3^2+j_3}} j^+  | j,m > \\[3mm]
| j,m-1> =  {\displaystyle 1 \over \displaystyle \sqrt{j(j+1)-m^2+m}}  j^-  | j,m >
={\displaystyle 1 \over \displaystyle \sqrt{\vec{j}^2 -j_3^2-j_3}} j^-  | j,m >
\end{array}
\label{qw7.15}
\ee

In the last step we replaced the numbers $s,m$ by operators.

Comparing (\ref{qw7.14}) with (\ref{qw7.11}) we find that
\ba
T^3&=&q^{-2j_3}[2j_3]={1 \over \lambda}(1-q^{-4j_3}) \nonumber \\[1mm]
T^+&=&{1 \over \lambda} q^{-(j_3+\hh)} \sqrt{{{\displaystyle q^{\sqrt{1+4\vec{j}^2}}+q^{-\sqrt{1+4\vec{j}^2}}
-q^{-2j_3+1}-q^{2j_3-1}} \over {\displaystyle \vec{j}^2 -j_3^2+j_3}}} j^+ 
\label{qw7.16} \\[2mm]
T^-&=&{1 \over \lambda} q^{-(j_3-\hh)} \sqrt{{{\displaystyle q^{\sqrt{1+4\vec{j}^2}}+q^{-\sqrt{1+4\vec{j}^2}}
-q^{-2j_3-1}-q^{2j_3+1}} \over {\displaystyle \vec{j}^2 -j_3^2-j_3}}} j^- \nonumber  \\[2mm]
\tau&=&q^{-4j_3} \nonumber
\ea 
have the same matrix elements (\ref{qw7.11}). That they satisfy the same algebra (\ref{qw7.1}) can be verified by a direct calculation. The eqns (\ref{qw7.13}) should be used in the
form
\[
(j_3-1)j^+=j^+j_3 \qquad \mbox{or} \qquad j_3j^+=j^+(j_3+1)
\] 
and
\[
j^+j^-=\vec{j}^2-j_3^2+j_3, \quad j^-j^+=\vec{j}^2-j_3^2-j_3
\]
We calculate:
\begin{eqnarray*}
T^+ T^-&=&{1 \over \lambda^2} q^{-2j_3+1} {{\displaystyle q^{\sqrt{1+4\vec{j}^2}}+q^{-\sqrt{1+4\vec{j}^2}}
-q^{-2j_3+1}-q^{2j_3-1}} \over {\displaystyle \vec{j}^2 -j_3^2+j_3}} j^+j^-  \\
T^- T^+&=&{1 \over \lambda^2} q^{-2j_3-1} {{\displaystyle q^{\sqrt{1+4\vec{j}^2}}+q^{-\sqrt{1+4\vec{j}^2}}
-q^{-2j_3+1}-q^{2j_3-1}} \over {\displaystyle \vec{j}^2 -j_3^2-j_3}} j^-j^+ \\
\hq T^+ T^--q  T^- T^+&=&{1 \over \lambda^2} q^{-2j_3} \left(q^{\sqrt{1+4\vec{j}^2}}+q^{-\sqrt{1+4\vec{j}^2}}-q^{-2j_3+1}-q^{2j_3-1}\right) \\
&=& {1 \over \lambda^2} q^{-2j_3} (q^{2j_3}-q^{-2j_3})(q-\hq) ={1 \over \lambda}(1-q^{-4j_3}) \\
&=&T^3
\end{eqnarray*}
This should serve as an example.

For $j = 1/2$ and $j = 1$ we show the representations explicitely:
\ba
T^+ | \hh,-\hh>&=&q^{-1} |\hh,\hh> \nn
T^+  | \hh,\hh>&=&0 \nn
T^- | \hh,-\hh>&=&0 \nn 
T^- | \hh, \hh>&=& q  | \hh,-\hh>
\label{qw7.17}\\
T^3  | \hh,-\hh>&=&-q  | \hh,-\hh> \nn
T^3  | \hh, \hh>&=&q^{-1} |\hh,\hh> \nonumber
\ea
\ba\label{qw7.18}
T^+ |1,-1>&=&\hq\sqrt{1+q^2} |1,0> \nn
T^+ |1,0>&=&{1 \over q^2} \sqrt{1+q^2} |1,1> \nn
T^+ |1,1>&=&0 \nn
T^-|1,-1>&=&0\nn
T^-|1,0>&=&q\sqrt{1+q^2}|1,-1>\\
T^-|1,1>&=&\sqrt{1+q^2}|1,0>\nn
T^3|1,-1>&=&-q(1+q^2)|1,-1>\nn
T^3|1,0>&=&0\nn
T^3|1,1>&=&\hq(1+\frac{1}{q^2})|1,1> \nonumber
\ea

We can identify the vectors of a representation with elements of a quantum
plane. The elements of a quantum plane can be multiplied. This product we
identify with the tensor product of the representations to obtain its 
transformation properties. From the comultiplication (\ref{qw7.5}) we find the rule how
the products transform:
\ba
T^3x\cdots=(\Delta(T^3) x \cdots)=(T^3 x) \cdots +(\tau x) T^3 \cdots \label{qw7.19}\\
T^{\pm}x\cdots=(\Delta(T^{\pm}) x \cdots)=(T^{\pm} x) \cdots +(\tau^{1/2}x) T^{\pm} \cdots \nonumber
\ea
the dots indicate the additional factors more explicitely for $j = 1/2$. We
identify
\be
|\hh,-\hh>=x^1, \quad | \hh,\hh>=x^2 
\label{qw7.20}
\ee
and obtain from (\ref{qw7.17})
\ba
T^3 x^1&=&q^2 x^1 T^3 -q x^1 \nn
T^3 x^2&=&q^{-2} x^2 T^3 +q^{-1} x^2 \nn
T^+ x^1&=&qx^1 T^+ +q^{-1} x^2 
\label{qwk7.21}\\
T^+ x^2&=&q^{-1} x^2 T^+ \nn
T^- x^1&=&q x^1 T^- \nn
T^- x^2&=&q^{-1} x^2 T^- +q x^1 \nonumber
\ea

Now we ask what are the algebraic relations on the variables $x^i$ that are
compatible with (3.18).

\begin{eqnarray*}
&T^+ x^1 x^2=(q x^1 T^+ +\hq x^2) x^2 =x^1 x^2 T^++\hq x^2 x^2 \\
&T^+ x^2 x^1=\hq  x^2 T^+ x^1=x^2 x^1 T^++\frac{1}{q^2} x^2 x^2 \\
&T^+ (x^1 x^2-q  x^2 x^1)=(x^1 x^2-q  x^2 x^1) T^+
\end{eqnarray*}

The relation
\be
x^1 x^2=q x^2 x^1
\label{qw7.22}
\ee
is compatible with \ref{qw3.18}. This can be verified for all the $T$s.

We have discovered the relation \ref{qw2.7} for the two-dimensional quantum plane.
This is not surprising, as we know that this plane is covariant under $SU_q(2)$
We could have started from \ref{qw2.7} and asked for all linear transformations of
the type (\ref{qw7.19}) that are compatible with \ref{qw2.7}. This way we would have found
 (\ref{qw7.19}) and from there the ``multiplication'' rule (\ref{qw7.1}) and the 
comultiplication (\ref{qw7.5}). It is all linked via the $\hat{R}$ matrix.

{From} \ref{qw4.1} and (\ref{qw7.9}) we can deduce the action of the $T$s on the conjugation
plane:  
 \ba
T^3 \overline{x}_1&=&\frac{1}{q^2} \overline{x}_1 T^3 +\hq \overline{x}_1 \nn
T^3 \overline{x}_2&=&q^2 \overline{x}_2 T^3 -q \overline{x}_2 \nn
T^+ \overline{x}_1&=& \hq \overline{x}_1 T^+ 
\label{qw7.23}\\
T^+ \overline{x}_2&=&q \overline{x}_2 T^+ - \overline{x}_1 \nn
T^- \overline{x}_1&=&\hq \overline{x}_1 T^- - \overline{x}_2 \nn
T^- \overline{x}_2&=&q \overline{x}_2 T^- \nonumber
\ea


\section{$q$-deformed Euclidean space in three dimensions}\label{sub2.7}

The $\hat{R}$-matrix for the fundamental representation of a quantum 
group contains all the information on the $\hat{R}$-matrices for the
other representations. One way to extract this information is to construct
quantum space variables as products of the quantum space variables of the
fundamental representation. Let us demonstrate this for $SU_q(2)$.

We start from the relations (\ref{qw2.11}) and the $\hat{R}$-matrix
(\ref{qw1.5}). We take four copies of the quantum planes $x,y,u$ and
$v$ such that

\ba
xu&=&\frac{1}{q} \hat R ux,\quad xy=\frac{1}{q}\hat Rvx\label{weqw9.1}\\
yu&=&\frac{1}{q} \hat R uy,\quad yv=\frac{1}{q}\hat Rvy\no
\ea
Then we consider ``bispinors'':

\be
xy\sim X,\quad uv \sim \widetilde X\label{weqw7.2}
\ee
We can use  (\ref{qwk7.21}) to single out those components that 
transform like a three-vector corresponding  to (\ref{qw7.18}):

\be
X^-=x^1y^1,\quad X^0=\frac{1}{\sqrt{1+q^2}}(x^1y^2+qx^2y^1),\quad X^+=x^2y^2
\ee
and the same for $\tilde X$ with $xy$ replaced by $uv$.

The relations (\ref{weqw7.2}) are sufficient to compute the $9\times 9$
$\hat{\cal R}$-matrix:

\be
X\widetilde X=\hat{\cal R} \widetilde X X
\ee

The relations are consistent with (\ref{qw7.9}) and the reality
condition on the quantum space:

\be
\overline{X^0}=X^0,\quad \overline{X^+}=-qX^-
\ee
This makes $X^0,X^+,X^-$ to be quantum space variable of
$SO_q(3)$.

The $\hat{\cal R}$-matrix will satisfy the Yang-Baxter equation if $\hat{R}$ 
satisfies it.

The $\hat{\cal R}$-matrix, calculated that way, has three different eigenvalues
$1(5),-q^{-4}(3)$ and $q^{-6}(1)$. The number in bracket is the multiplicity of the
respective eigenspaces and corresponds to the $j=2,\,j=1$
and $j=0$ representations.

As a consequence, $\hat{\cal R}$ satisfies the characteristic equation:

\be
(\hat{\cal R} -1)(\hat{\cal R} +\frac{1}{q^4})(\hat{\cal R} -\frac{1}{q^6})=0
\ee
which in turn gives the projectors on the irreducible subspaces:

\ba
P_1&=&\frac{q^{12}}{(1+q^2)(1-q^6)}(\hat{\cal R} -1)(\hat{\cal R}+\frac{1}{q^6})\no\\
P_3&=&\frac{q^{10}}{(1+q^2)(1+q^4)}(\hat{\cal R} -1)(\hat{\cal R}-\frac{1}{q^4})\\
P_5&=&\frac{q^{10}}{(q^4+1)(q^6-1)}(\hat{\cal R} +\frac{1}{q^4})(\hat{\cal R}-\frac{1}{q^6})\no
\ea

The projectors contain the Clebsch-Gordon coefficients for the 
reduction of the $3\otimes 3$ representation of $SO_q(3)$. Thus the Clebsch-Gordon
coefficients can be derived from the $\hat{\cal R}$-matrix for $q\ne 1$. For $q=1$
the eigenvalues degenerate and it is not possible to
obtain all the projectors from $\hat{\cal R}$. But we could compute the 
projectors first and then put $q=1$.

The projector $P_1$ projects on a one-dimensional subspace, we can
define a metric from this projection:
\ba
\widetilde X \circ X&=&\widetilde X^0X^0-q\widetilde X^+X^--\frac{1}{q}\widetilde X^-X^+=\widetilde X^AX^B\eta_{BA}\\
\eta_{00}&=&1\,,\quad \eta_{+-}=-\frac{1}{q}\,,\quad \eta_{-+}=-q\no
\ea
With the metric $\eta_{AB}$ we can lower vector indices:
\be
X_B=X^A\eta_{AB},
\ee
We raise them with $\eta^{AB}$
\be
X^B=\eta^{BC}X_C 
\ee
This defines $\eta^{BC}$ by the equation

\be
\eta^{BA}\eta_{BC}=\delta^A_C,\quad \eta^{AB}\eta_{CB}=\delta^A_C\label{weqw7.11} 
\ee
The projector $P_1$ in terms of the metric is

\be
(P_1)_{CD}^{AB}=\frac{q^2}{1+q^2+q^4}\eta^{AB}\eta_{DC}\label{weqw7.12}
\ee

In a similar way we can use the Clebsch-Gordon coefficients that 
combine two vectors to a vector for the definition of a $q$-deformed
$\varepsilon$-tensor:

\be
Z^A=\widetilde X^CX^B\varepsilon_{BC}{}^A\label{weqw7.13}
\ee
We find for the non-vanishing components:

\ba
\varepsilon_{+-}{}^0&=&q,\quad\varepsilon_{-+}{}^0=-q,\quad\varepsilon_{00}{}^0=1-q^2\no\\
\varepsilon_{+0}{}^+&=&1,\quad\varepsilon_{0+}{}^+=-q^2\label{weqw7.14}\\
\varepsilon_{-0}{}^-&=&-q^2,\quad\varepsilon_{0-}{}^-=1\no
\ea
The indices of the $\varepsilon$ tensor can be raised and lowered with
the metric:

\be
\varepsilon_{ABC}=\varepsilon_{AB}{}^D\eta_{DC}\label{weqw7.15}
\ee
The projector $P_3$ in terms of the $\varepsilon$ tensor is:

\be
{P_3}_{CD}^{AB}=\frac{1}{1+q^4}\varepsilon^{FAB}\varepsilon_{FDC}\label{weqw7.16}
\ee
The projector $P_5$ can be obtained from the relation

\be
P_1+P_3+P_5=1\label{weqw7.17}
\ee
The $\hat{\cal R}$-matrix is a sum of projectors as well.

\be
\hat{\cal R}=P_5-\frac{1}{q^4}P_3+\frac{1}{q^6}P_1\label{weqw7.18}
\ee

We can ask for the most general linear combination of the projectors
that solves the Yang-Baxter equation and obtain
$\hat{\cal R}$ and $\hat{\cal R}^{-1}$.

A natural way to define the $3$-dimensional Euclidean quantum space is:

\ba
\varepsilon_{FDC}X^CX^D&=&0\no\\
X^0X^+&=&q^2X^+X^0\label{weqw7.19}\\
X^-X^0&=&q^2X^0X^-\no\\
X^-X^+&=&X^+X^-+(q-\frac{1}{q})X^0X^0\no
\ea

Derivatives can be defined along the line of chapter 3:

\ba
\partial_BX^A&=&\delta_B^A+q^4\hat{\cal R}_{BD}^{AC}X^D\partial_C\label{weqw7.20}\\
\varepsilon^{FBA}\partial_A\partial_B&=&0\no
\ea

To summarize we have constructed an algebra freely generated by 
elements of the quantum space $x$ and derivative $\partial$, this algebra
is divided by the ideal generated by the relations (\ref{weqw7.19}) and
(\ref{weqw7.20}). We have constructed it in such a way that the 
Poincar{\'e}-Birkhoff-Witt property holds and that the algebra allows
the action of $SO_q(3)$.

For the definition of a conjugation that is consistent with (\ref{weqw7.20})
we first extend the algebra by enlarging it by conjugate derivatives.
We use the notation:

\be
\overline{\partial^A}=\overline \partial_A\label{weqw7.21}
\ee

and obtain from (\ref{weqw7.20})
and (\ref{qw7.5})
\ba
\overline \partial_CX_D&=&-\frac{1}{q^6}\eta_{CD}+\hat{\cal R}^{BA}_{DC}X_A\partial_B\label{weqw7.22}\\
\varepsilon_{FAB}\overline \partial^B\overline \partial^A&=&0\no
\ea

The $\partial,\,\overline\partial$ relations are not yet specified. We can apply $\partial\overline\partial$ and $\overline\partial\partial$ to 
$X$ as we did in (\ref{qw5.7}) and find the consistent relation:
\be
\varepsilon_{FAB}(\partial^B\overline\partial^A+\overline\partial^B\partial^A)=0\label{weqw7.23}
\ee

The $X,\,\partial,\,\overline\partial$ algebra divided by (\ref{weqw7.19}), (\ref{weqw7.20})
and (\ref{weqw7.22}) still satisfies Poincar{\'e}-Birkhoff-Witt, is
$SO_q(3)$-covariant and consistent with (\ref{qw7.5}).

Miraculously it turns out that $\overline\partial$ can again be related non-linearly
to $\partial$ as it was the case in (\ref{qw5.11}). This can be achieved by the 
scaling operator $\Lambda$:
\ba
\Lambda&=&q^6\{1+(q^4-1)X\circ\partial+q^2(q^2-1)^2(X\circ X)(\partial\circ\partial)\}\label{weqw7.24}\\
\Lambda X^A&=&q^4X^A\Lambda\label{weqw7.25}\\
\Lambda \partial^A&=&q^{-4}\partial^A\Lambda\no
\ea

If we now define an operator

\be
\overline\partial^A=-\Lambda^{-1}\left(\partial^A+q^2(q^2-1)X^A(\partial\circ\partial)\right)\label{weqw7.26}
\ee
we find that this operator satisfies the relations (\ref{weqw7.22}) 
and (\ref{weqw7.23}). Thus we divide our algebra once more by the 
ideal generated by  (\ref{weqw7.26}). This is exactly the same
strategy we used in chapter 5. From  (\ref{weqw7.26}) follows

\be
\overline \Lambda=\Lambda^{-1}\label{weqw7.27}
\ee

As in (\ref{qw5.13}) we define the momenta

\be
P^A=-\frac{i}{2}(\partial^A-\overline\partial^A)\label{weqw7.28}
\ee

Now we are in a position to derive the $q$-deformed Heisenberg
algebra for the three-dimensional quantum space defined by 
(\ref{qw7.5}) and (\ref{weqw7.19}):

\ba
\varepsilon_{FAB}P^AP^B&=&0\label{weqw7.29}\\
P^AX^B-(\hat{\cal R}^{-1})^{AB}_{CD}X^CP^D&=&-\frac{i}{2}q^3\Lambda^{-\frac{1}{2}}\left\{(1+\frac{1}{q^6})\eta^{AB}W-(1-\frac{1}{q^4})\varepsilon^{ABF}L_F\right\}
\no
\ea

The operators $\Lambda$, $W$ and $L^A$ at the right hand side of eqn
(\ref{weqw7.29}) are defined as differential operators, eqn. (\ref{qw5.8})
should serve as an example. The separation of the various terms has been
done by transformation properties (singlet, triplet) and by 
factoring the differential operators into a product of a unitary
and a hermitean operator. This is how $W$, $L_A$ and $\Lambda$ 
have been defined.

\be
\overline\Lambda=\Lambda^{-1},\quad\overline W=W \quad\mbox{and}\quad\overline{L^A}=L_A
\label{weqw7.30}
\ee

It turns out that the differential operators form an algebra by
themselves
\ba
\varepsilon_{BA}{}^CL^AL^B&=&-\frac{1}{q^2}WL^C\no\\
L^AW&=&WL^A,\label{weqw7.31}\\
\Lambda L^A&=&L^A \Lambda,\quad\Lambda W=W\Lambda
\no
\ea
and in addition

\be
W^2-1=q^4(q^2-1)^2L\circ L\label{weqw7.32}
\ee

We can now take the relations (\ref{weqw7.31}) and (\ref{weqw7.32}) as
the defining relations for the $\Lambda,W,L_A$ algebra and
consider $X^A$ and $P^A$ as models under this algebra.

In this way we arrive at the algebra that generalizes the 
Heisenberg algebra to a $SO_q(3)$ structure.

We summarize the algebra:

q-deformed phase space:
\ba
X^CX^D\varepsilon_{DC}{}^B&=&0,\quad\overline{X^A}=X_A\label{weqw7.33}\\
P^CP^D\varepsilon_{DC}{}^B&=&0,\quad\overline{P^A}=P_A\no
\ea

$SO_q(3)$:
\ba
L^CL^B\varepsilon_{BC}{}^A&=&-\frac{1}{q^2}WL^A,\quad \overline{L^A}=L_A,\quad
L^AW=WL^A,\quad \overline W=W\label{weqw7.34}\\
W^2-1&=&q^4(q^2-1)^2L\circ L\no
\ea

scaling operator:
\ba
\Lambda^{\frac{1}{2}}L^A&=&L^A\Lambda^{\frac{1}{2}}\quad \overline{\Lambda^{\frac{1}{2}}}=\Lambda^{-\frac{1}{2}}\label{weqw7.35}\\
\Lambda^{\frac{1}{2}}W&=&W\Lambda^{\frac{1}{2}}\no
\ea

comodule relations for $SO_q(3)$:

\ba
L^AX^B&=&-\frac{1}{q^4}\varepsilon^{ABC}X_CW-\frac{1}{q^2}\varepsilon_{KC}{}^A\varepsilon^{KBD}L_D\label{weqw7.36}\\
WX^A&=&(q^2-1+\frac{1}{q^2})X^AW+(q^2-1)^2\varepsilon^{ABC}X_CL_B
\no
\ea

The coordinates can be replaced by the momenta to obtain the module 
relations for $P_A$.

Scaling properties:
\ba
\Lambda^{\frac{1}{2}}X^A=q^2X^A\Lambda^{\frac{1}{2}}\label{weqw7.37}\\
\Lambda^{\frac{1}{2}}P^A=q^{-2}P^A\Lambda^{\frac{1}{2}}\no
\ea

generalized (mimicked) Heisenberg relation:

\be
P^AX^B-(\hat{\cal R}^{-1})_{CD}^{AB}X^CP^D=-\frac{i}{2}q^3\Lambda^{-\frac{1}{2}}\left\{(1+\frac{1}{q^6})\eta^{AB}W-(1-\frac{1}{q^4})\varepsilon^{ABF}L_F\right\}\label{weqw7.38}
\ee

Eqns. (\ref{weqw7.33}) to (\ref{weqw7.38}) are the defining relations for
the algebra we will be concerned with. Part III of this
lecture should analyze this algebra in the same way as we analyzed the
one-dimensional Heisenberg algebra in part I. The representation 
theory of the new algebra has been thoroughly investigated, it leads to
very similar phenomena as we saw in part I for the representations.
The $q$-deformed Minkowski structure has been analyzed in the same way.

\end{document}